\documentclass[12pt,english,onecolumn, draftcls, conference]{IEEEtran}

\usepackage[T1]{fontenc}
\usepackage[latin1]{inputenc}
\usepackage{geometry}
\usepackage{amsmath,amssymb,amsthm,mathrsfs,amsfonts,dsfont}
\usepackage{graphicx}
\usepackage{cite}
\usepackage{color}
\usepackage{mathtools}
\usepackage{cases}
\usepackage{babel}
\usepackage{soul}
\usepackage{accents}
\usepackage{babel}

\allowdisplaybreaks
\newtheorem{rem}{Remark}[section]
\theoremstyle{plain}
\newcommand*{\Scale}[2][4]{\scalebox{#1}{$#2$}}

\begin{document}
\title{Zero-Delay Joint Source-Channel Coding in the Presence of Interference Known at the Encoder}
\author{Morteza Varasteh$^\dagger$, Deniz G\"{u}nd\"{u}z$^\dagger$, and Ertem Tuncel$^*$\\
$^\dagger$ Department of Electrical and Electronic Engineering, Imperial College London, London, U.K.\\
$^*$ Department of Electrical and Computer Engineering, University of California, Riverside, CA.\\
Email: m.varasteh12@imperial.ac.uk, d.gunduz@imperial.ac.uk, ertem.tuncel@ucr.edu}

\maketitle
\begin{abstract}

Zero-delay transmission of a Gaussian source over an additive white Gaussian noise (AWGN) channel is considered in the presence of an additive Gaussian interference signal. The mean squared error (MSE) distortion is minimized under an average power constraint assuming that the interference signal is known at the transmitter. Optimality of simple linear transmission does not hold in this setting due to the presence of the known interference signal. While the optimal encoder-decoder pair remains an open problem, various non-linear transmission schemes are proposed in this paper. In particular, interference concentration (ICO) and one-dimensional lattice (1DL) strategies, using both uniform and non-uniform quantization of the interference signal, are studied. It is shown that, in contrast to typical scalar quantization of Gaussian sources, a non-uniform quantizer, whose quantization intervals become smaller as we go further from zero, improves the performance. Given that the optimal decoder is the minimum MSE (MMSE) estimator, a necessary condition for the optimality of the encoder is derived, and the numerically optimized encoder (NOE) satisfying this condition is obtained. Based on the numerical results, it is shown that 1DL with non-uniform quantization performs closer (compared to the other schemes) to the numerically optimized encoder while requiring significantly lower complexity.
\end{abstract}
\global\long\def\thefootnote{\arabic{footnote}}
\setcounter{footnote}{0}

\section{Introduction}
While the spectral efficiency of communication systems has improved significantly within the last decade, latency remains as the bottleneck for many applications. In many emerging applications, such as those involving \emph{cyber-physical systems} (CPS) or \emph{wireless sensor networks} (WSN), real-time interaction among distributed autonomous agents is crucial. A communication link is called real-time when the communication time is lower than the time constants of the application. Such applications impose significantly lower round-trip latency requirements compared to what is achievable today. For example, in many applications involving CPSs, local system measurements are reported by sensor nodes via noisy links to other network agents. The need to have near real-time monitoring and control of the underlying physical system imposes strict delay constraints on the communication links. In such a scenario, utilizing long block codes for source compression or channel coding is not viable due to the stringent delay constraint. Similarly, when tactile control of an object and hearing/ seeing its reaction through a wireless connection is desired, a reaction latency on the order of milliseconds will be imposed on the communication link \cite{Fettweis}. For example, for a typical 1 m/s speed of a finger on a touch screen, the reaction time for the screen is expected to be approximately 1ms in order to achieve an unnoticeable displacement of 1mm between the object to be moved and the finger \cite{Fettweis-Alamouti}.

We consider zero-delay transmission of system parameters over wireless channels, that is, a single source sample needs to be transmitted over a single use of the channel. It is well-known that zero-delay linear encoding (uncoded transmission) of a Gaussian source over an \emph{additive white Gaussian noise} (AWGN) channel does not result in any performance loss in terms of the end-to-end \emph{mean-squared error} (MSE) distortion \cite{goblick1965theoretical}. However, this is not the case if there is bandwidth mismatch between the source and channel \cite{Akyol_Viswanatha_Rose_2012,Kotel'nikov_1959,hekland_floor_ramstad_2009}, if there is correlated source side information at the receiver \cite{Xuechen_Tuncel_2011}, or if there is a peak power constraint at the transmitter \cite{Gunduz_inaki_varasteh_2013}. Characterization of the optimal transmission strategy is challenging in general, and remains an open problem in most cases.

\begin{figure}
\begin{centering}
\includegraphics[scale=0.8]{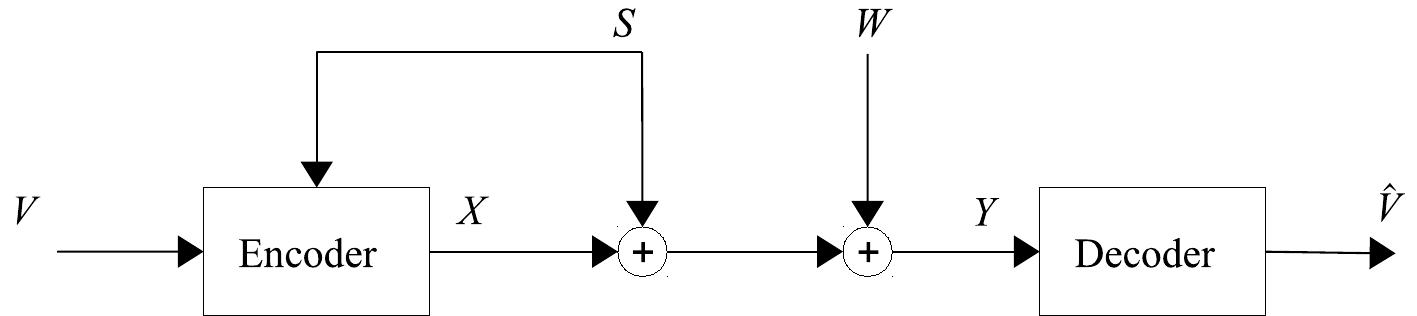}
\par\end{centering}
\caption{\label{Fig:AWGNInterference}Zero-delay transmission of a Gaussian source over an AWGN channel in the presence of AWG interference known at the transmitter.}
\vspace{0mm}
\end{figure}

In this paper we consider zero-delay transmission of a Gaussian source over an AWGN channel in the presence of an \emph{additive white Gaussian} (AWG) interference signal causally known at the transmitter. This is known as the \textit{dirty-tape channel}. Known interference at the transmitter can be used to model communication systems which use superposition coding to transmit multiple data streams simultaneously \cite{Cover:1972,gunduz2008joint}. For the superposed data streams, the codewords corresponding to lower layers act as known interference. The capacity of the dirty-tape channel was first studied by Shannon \cite{Shannon-side-information}, who characterized the capacity using the so-called \emph{Shannon strategies}. The channel model when the interference is known non-causally at the transmitter is known as the \textit{dirty-paper channel}. The capacity of the dirty-paper channel was characterized by Gelfand and Pinsker in \cite{Gelfand_1980}, and it was later shown in \cite{costa1983} that, in the Gaussian setting, the capacity of the dirty-paper channel is equal to the one without interference.

Despite Shannon's single-letter characterization, there is no closed-form capacity expression for the dirty-tape channel even in the Gaussian setting. Willems in \cite{willems_2000} proposed the \emph{interference concentration (ICO)} strategy for the Gaussian dirty-tape channel. The basic idea of this scheme is cancelling the interference by giving a structure to it. Willems showed that, it is possible to partially cancel the interference at the receiver by quantizing it at the encoder, and by proper power allocation between the interference quantization error and the channel input signal, which is uniformly distributed over the quantization region. More recently, Erez \emph{et al}. \cite{Erez_shamai_zamir_2005} proposed inflated lattice strategies for the Gaussian dirty-tape channel in \cite{willems_2000}. They show that the rate loss of their coding scheme with respect to no interference, which is shown to be zero in the case of dirty-paper channel \cite{costa1983}, is not more than 0.254 bits per channel use in the asymptotic high \emph{signal to noise ratio} (SNR) regime. On the other hand, it is shown in \cite{Khosravi_Farasani_akhbari_aref_2010} that the ICO scheme of Willems performs better than the inflated lattice based coding scheme in the low SNR regime. In \cite{Skoglun_Larsson_2008}, optimal mappings based on an iterative algorithm are proposed. Based on the numerical results in \cite{Skoglun_Larsson_2008}, it is shown that the numerically obtained encoder performs well compared to the scheme proposed in \cite{Erez_shamai_zamir_2005}.

All of the above mentioned work study the channel coding problem whereas we are interested in zero-delay \emph{joint source channel coding} (JSCC) over the dirty-tape channel. Note that Shannon's source-channel separation theorem \cite{Cover:book} does not apply to zero-delay JSCC problems; and hence, we can not directly use the above channel coding results to evaluate the MSE performance. A generalization of this problem is studied in \cite{saleh_alajaji_2014}, which further allows correlation between the source and the interference signals, and bandwidth mismatch between the source and the channel. While \cite{saleh_alajaji_2014} focuses on deriving a numerically optimized encoder and decoder pair, our goal here is to develop low-complexity joint source-channel transmission techniques motivated by the channel coding strategies proposed in \cite{willems_2000} and \cite{Erez_shamai_zamir_2005}. Expanding upon our previous work in \cite{varasteh_gunduz_tuncel}, we consider ICO and \emph{one-dimensional lattice} (1DL) schemes combined with nonlinear companders. While characterizing the optimal performance is elusive for this problem, we present numerical results comparing the performance of the proposed strategies, and provide some heuristics to improve them. In particular, we propose a counter-intuitive non-uniform quantization scheme in conjunction with the ICO and 1DL schemes, which increases the average quantization error, and hence, the power used for interference concentration, but leads to a lower MSE since the transmitter can then use a compander with a larger dynamical range for the more likely interference states.

Similarly to \cite{saleh_alajaji_2014}, we also characterize the  necessary condition for the optimality of an encoder mapping, and obtain a \emph{numerically optimized encoder} (NOE) using steepest descent to search for an encoder that satisfies the derived necessary condition. While the MSE achieved by NOE outperforms the other proposed schemes, it is demanding computationally. It is shown that non-uniform quantization in conjunction with 1DL performs closer (compared to the other schemes) to the NOE, while the number of parameters to be optimized for the 1DL scheme (with non-uniform quantizer) is significantly less than NOE; and hence, it has significantly less computational complexity.

The rest of the paper is organized as follows: In Section \ref{Sec:system_model} we introduce the system model. In Section \ref{Sec:achievable transmission schemes} zero-delay transmission schemes under average power constraint are introduced. In Section \ref{Sec:Numerically Optimized encoder}, we characterize the necessary condition for the optimal encoder, and introduce NOE. In Section \ref{Sec:simulation}, we compare all the proposed transmission schemes numerically, and in Section \ref{Sec:conclusion} we conclude the paper.

\section{system model}\label{Sec:system_model}
We consider the transmission of a Gaussian source over an AWGN channel in the presence of an AWG interference signal, which is known at the transmitter. The setup is illustrated in Fig. \ref{Fig:AWGNInterference}. Without loss of generality, we assume, that the memoryless Gaussian source sample, $V$, has zero mean and unit variance, i.e., $V\sim\mathcal{N}(0,1)$. The interference signal is independent of the source, and also follows a Gaussian distribution, $S\sim\mathcal{N}(0,\sigma_s^2)$. The discrete memoryless channel output $Y$, is given by $Y=X+S+W$, where $X$ is the channel input, $S$ is the known Gaussian interference signal, and $W$ is the additive Gaussian noise, $W\sim\mathcal{N}(0,\sigma_n^2)$, independent of the source and interference signals.

We denote the zero-delay encoding function as $X=h(V,S)$. An average power constraint is imposed on the channel input:
\begin{equation}\label{Eq:power_constraint}
\mathbb{E}[X^2] \leq P,
\end{equation}
where the expectation is over all realizations of the source and interference signal. We are interested in transmitting the source samples, $V$, over the channel under MMSE criterion. We denote the MMSE estimation function at the receiver by $\hat{V}=g(Y)\triangleq\mathbb{E}[V|Y]$. Our goal is to characterize the minimum MSE $\mathbb{E}[|V-\hat{V}|^2]$, for given $P$, $\sigma_s^2$ and $\sigma_n^2$ values.

We note that, in our setting, due to the zero-delay constraint, causal and non-causal knowledge of the interference are equivalent. In other words, non-causal knowledge of the interference is useless, and the transmitter only uses the knowledge of the current value of the interference.

We define the functions below, which will be used throughout the paper.
\begin{align}\nonumber
\mathcal{I}_0(m_1,m_2,a,b)&\triangleq\frac{\sqrt{m_1}e^{-\frac{m_2^2}{4m_1}}}{\sqrt{\pi}}\int\limits_{a}^{b}e^{-m_1 u^2 +m_2 u}du\\\nonumber
&=\text{Q}\left(\frac{m_2-2bm_1}{\sqrt{2m_1}}\right)-\text{Q}\left(\frac{m_2-2am_1}{\sqrt{2m_1}}\right),\\\nonumber
\mathcal{I}_1(m_1,m_2,a,b)&\triangleq\int\limits_{a}^{b}u e^{-m_1 u^2 +m_2 u}du\\\nonumber
&=\frac{1}{2m_1}\left(e^{-m_1 a^2+m_2 a}-e^{-m_1 b^2+m_2 b}\right)+\frac{\sqrt{\pi}m_2e^{\frac{m_2^2}{4m_1}}}{2m_1\sqrt{m_1}}\cdot
\mathcal{I}_0(m_1,m_2,a,b),\\\nonumber
\mathcal{I}_2(m_1,m_2,a,b)&\triangleq\int\limits_{a}^{b}u^2e^{-m_1 u^2 +m_2 u}du\\\nonumber
&=\frac{m_2}{4m_1^2}\left(e^{-m_1 a^2+m_2 a}-e^{-m_1 b^2+m_2 b}\right)+ \frac{1}{2m_1}\left(ae^{-m_1 a^2+m_2 a}-be^{-m_1 b^2+m_2 b}\right)\\\nonumber
&+\frac{\sqrt{\pi}e^{\frac{m_2^2}{4m_1}}}{2m_1\sqrt{m_1}}\cdot\left(1+\frac{m_2^2}{2m_1}\right)\cdot
\mathcal{I}_0(m_1,m_2,a,b),
\end{align}
where $\text{Q}(\cdot)$ is the complementary cumulative function and is defined as $\text{Q}(t)=\int\limits_{t}^{\infty}\frac{1}{\sqrt{2\pi}}e^{-\frac{t^2}{2}}$. Since we deal with definite integrals throughout the paper, we will avoid writing the boundaries of the integrals explicitly when they are from $-\infty$ to $\infty$. Also, if no limits are specified, the summations are over all integers $\mathbb{Z}$. We also define the rectangle function $R(t)$ as
\begin{align}\nonumber
R(t)&=\left\{\begin{array}{ccl}
                  1 &\text{if}& \quad -\frac{1}{2}\leq t \leq \frac{1}{2} \\
                  0 &\text{if}& \quad \text{otherwise}
                \end{array}
\right..
\end{align}

\section{ Parameterized zero-delay transmission schemes}\label{Sec:achievable transmission schemes}
In this section we introduce five different transmission schemes for the setup introduced in Section \ref{Sec:system_model} with increasing complexity. Later on, in Section \ref{Sec:simulation}, we will compare and comment on the performances of these schemes.

\subsection{Interference Cancellation (ICA)}\label{Sub:ICA}

The simplest way to communicate in the presence of a known interference signal is to cancel the interference. In the \textit{interference cancellation} (ICA) scheme, the transmitted signal $X$ is a simple linear combination of the source realization $V$ and the interference $S$. The transmitter decides how much of the interference will be cancelled depending on the system parameters. We have
\begin{align}\nonumber
X=aV+bS,
\end{align}
where $a$ and $b$ are the coefficients to be determined. The channel input has to satisfy
\begin{align}\label{Eq:power_constraint}
\mathbb{E}[X^2]=a^2+b^2 \sigma_s^2\leq P.
\end{align}

With MMSE estimation at the receiver, the achievable average distortion is found as
\begin{equation}\label{Eq:D_ICA}
D_{ICA}=\frac{1}{1+\frac{P-b^2\sigma_s^2}{(b+1)^2\sigma_s^2+\sigma_n^2}}.
\end{equation}

The optimal b value that minimizes (\ref{Eq:D_ICA}) is given by
\begin{equation}\label{Eq:b_star}
%b^*=-\frac{P+\sigma_s^2+\sigma_n^2 - \sqrt{(P+\sigma_n^2)^2+\sigma_s^4-2\sigma_s^2(P-\sigma_n^2)}}{2\sigma_s^2}.
b^*=-\frac{P+\sigma_s^2+\sigma_n^2 - \sqrt{(P-\sigma_s^2)^2+\sigma_n^2+2\sigma_n^2(P+\sigma_s^2)}}{2\sigma_s^2}.
\end{equation}

The optimal value for $a$ can be obtained from (\ref{Eq:power_constraint}) and (\ref{Eq:b_star}). The ICA scheme consumes part of the transmission power for interference cancellation; and thus, is expected to perform poorly especially in the low power regime, when the interference power is relatively high compared to the input power.

\begin{rem}
We note that in the high signal to interference and noise ratio (SINR) regime ($P\gg\sigma_s^2$) the average achievable distortion is $D_{\text{ICA}}=\frac{1}{1+\frac{P-\sigma_s^2}{\sigma_n^2}}$. That is because, by rewriting $b^*$ we have
\begin{align}
b^{*}&=-\frac{P+\sigma_s^2+\sigma_n^2-\sqrt{P^2(1-\frac{2\sigma_s^2}{P}+\frac{2\sigma_n^2}{P}+\frac{\sigma_s^4+\sigma_n^2+2\sigma_n^2\sigma_s^2}{P^2})}}{2\sigma_s^2}\\
&\stackrel{\mathclap{\mbox{(\textsl{a})}}}{\simeq} -\frac{P+\sigma_s^2+\sigma_n^2-P(1-\frac{\sigma_s^2}{P}+\frac{\sigma_n^2}{P})}{2\sigma_s^2}\\
&=-1,
\end{align}
%\stackrel{\mathclap{\mbox{(a)}}}{\simeq}
where (\textsl{a}) is due to the approximation $\sqrt{1-x}\simeq 1-x/2$ for small $x$. This is as though the signal $X=aV-S$, $(a=P-\sigma_s^2)$ is transmitted over the channel.
\end{rem}

The analysis of the performance of the ICA scheme is relegated to Section \ref{Sec:simulation}. Next  we will introduce alternative non-linear transmission strategies.

\subsection{Interference Concentration (ICO)}\label{Sec:ICO}

This scheme is motivated by Willems' ICO scheme for channel coding \cite{willems_2000}. We combine the interference concentration idea with JSCC of a Gaussian signal under a \emph{peak power constraint} (PPC) \cite{Gunduz_inaki_varasteh_2013}. The interference signal $S$ is concentrated to one of the pre-determined discrete points on the real line; that is, the interference is quantized, and the corresponding quantization noise is cancelled, rather than cancelling the whole interference. Only the signal corresponding to the quantization index of the interference is received at the receiver. The transmitter superposes a companded version of the source signal such that it is compressed into one quantization interval of the quantizer.

\begin{figure}
\begin{centering}
\includegraphics[scale=0.7]{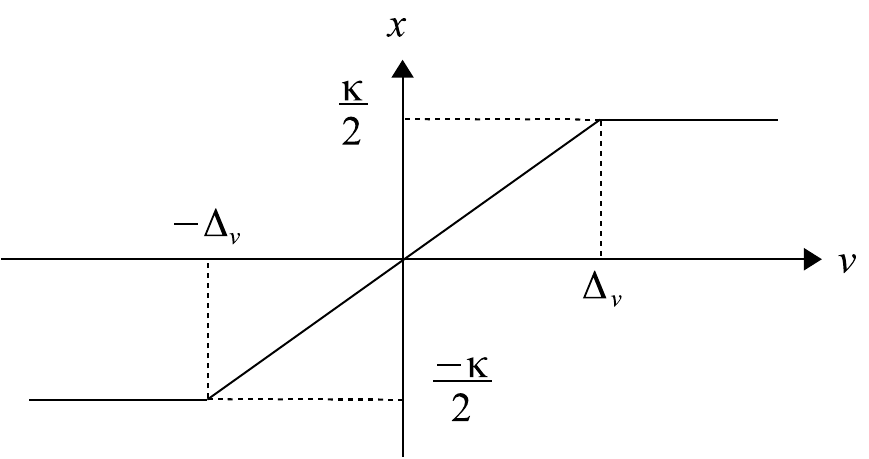}
\par\end{centering}
\caption{\label{Fig_figure9}Source clipping and mapping at the transmitter.}
\vspace{0mm}
\end{figure}

The signal transmitted over the channel is given by
\begin{equation}\label{Eq:channel input}
X=T(V)- (S~\text{mod}~\Delta),
\end{equation}
where $(S~\text{mod}~\Delta) \in[-\frac{\Delta}{2},\frac{\Delta}{2})$ corresponds to the quantization error, and is defined as
\begin{equation}\label{mod}
S~\text{mod}~\Delta\triangleq S-\mathcal{Q}(S),
\end{equation}
where $\mathcal{Q}(S)$ is the nearest neighbour quantizer defined as below
\begin{equation}\label{Q}
\mathcal{Q}(S)\triangleq \Delta \cdot \left\lfloor\frac{S}{\Delta} + \frac{1}{2}\right\rfloor,
\end{equation}
where $\lfloor \cdot \rfloor$ is the floor operation.

\begin{figure}
\begin{centering}
\includegraphics[scale=0.3]{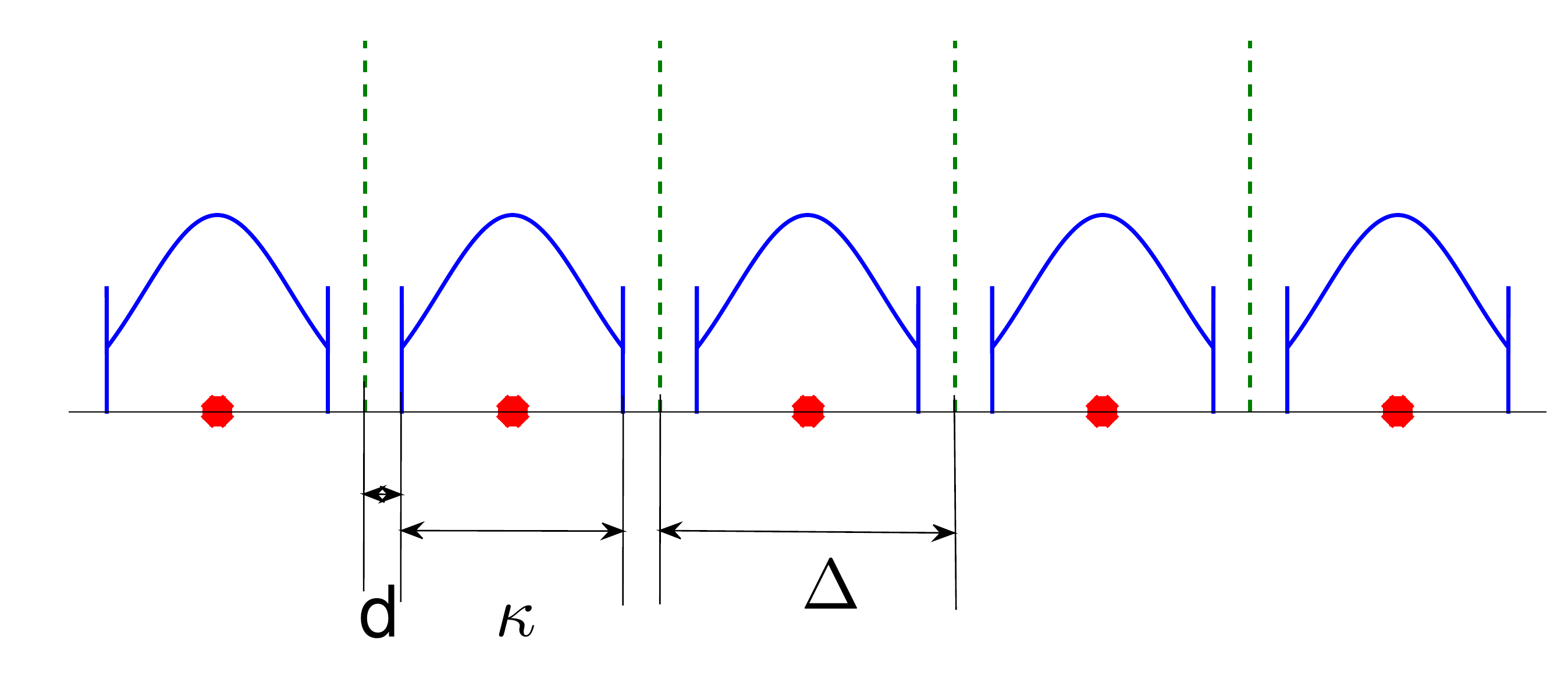}
\par\end{centering}
\caption{\label{Fig_figure0}Source is clipped to the region $[-\frac{\kappa}{2},\frac{\kappa}{2}]$, and the channel input is limited to the interval $[-\frac{\kappa+\Delta}{2},\frac{\kappa+\Delta}{2}]$. Dots are interference concentration points and dashed lines are the decision thresholds for interference concentration at the transmitter.}
\vspace{0mm}
\end{figure}

The source is clipped and mapped as in Fig. \ref{Fig_figure9} to the interval $[-\kappa/2,\kappa/2]$ as below
\begin{align}
\label{eq:scaling}
T(v)&=\left\{ \begin{array}{ccc}
           \frac{\kappa}{2} ,&\quad v\geq \frac{\Delta_v}{2}, \\
           \frac{\kappa}{\Delta_v} v,&\quad -\frac{\Delta_v}{2}\leq v<\frac{\Delta_v}{2},\\
           -\frac{\kappa}{2} ,&\quad v< -\frac{\Delta_v}{2}.
         \end{array}\right.
\end{align}
where $\kappa=\Delta-2d$. Notice that $\kappa+\Delta$ is the variation range of the channel input $X$ (since $T(v)$ and $(S~\text{mod}~\Delta)$ are varying in the intervals $[-\kappa/2,\kappa/2]$, and $[-\Delta/2,\Delta/2]$), respectively. Parameter $d$ can be considered as a guard interval between the source mappings into different intervals. Parameters $d,~\kappa$ and $\Delta$ are illustrated in Fig. \ref{Fig_figure0}.

It can be seen from (\ref{Eq:channel input}) that, in the ICO scheme, $s$ is concentrated to one of the quantization indices in $\{i\Delta:i\in \mathbb{Z}\}$, which corresponds to a uniform quantizer with quantization interval size of $\Delta$. Power consumed by the transmitter for interference concentration is equivalent to the average quantization noise variance for the interference signal. While the power allocated to interference concentration, $\sigma_{S \text{mod} \Delta}^2$, depends only on the value of $\Delta$, the power of the compander component, $\sigma_{T}^2$, depends on $\kappa$ and $\Delta_v$ parameters. $\sigma_{T}^2$ and $\sigma_{S \text{mod} \Delta}^2$ are to be chosen such that the channel power constraint is satisfied. We have
\begin{equation}\label{Eq:1}
    \mathbb{E}[X^2]=\sigma_{T}^2+\sigma_{S \text{mod} \Delta}^2\leq P,
\end{equation}
where the expectation is taken over the \emph{probability density function} (pdf) of the channel input $f_X(x)=f_T(t)\star f_{S\text{mod}\Delta}(t)$, where $\star$ denotes the convolution operation, and we have
\begin{align}\nonumber
f_{T}(t) &= \frac{1}{\sqrt{2\pi}\alpha} e^{-\frac{t^2}{2\alpha^2}} R\left(\frac{t}{\kappa}\right) + \text{Q}\left(\frac{\Delta_v}{2}\right)\cdot \left(\delta\left(t-\frac{\kappa}{2}\right)+\delta\left(t+\frac{\kappa}{2}\right)\right),\\\nonumber
f_{S\text{mod}\Delta}(t)&=\sum\limits_{i}\frac{1}{\sqrt{2\pi}\sigma_s}e^{-\left(\frac{i\Delta+t}{\sqrt{2}\sigma_s}\right)^2}\cdot R\left(\frac{t}{\Delta}\right).
\end{align}
Therefore, for $\sigma_{T}^2$ and $\sigma_{S \text{mod} \Delta}^2$ we have
\begin{align}\label{Eq:14}
\sigma_{T}^2&= \kappa^2\left[\frac{1}{\Delta_v^2} +\text{Q}\left(\frac{\Delta_v}{2}\right)\left(\frac{1}{2}-\frac{2}{\Delta_v^2}\right)-\frac{e^{-\frac{\Delta_v^2}{8}}}{\Delta_v\sqrt{2\pi}}\right],\\\label{Eq:15}
\sigma_{S \text{mod} \Delta}^2&=\frac{1}{\sqrt{2\pi}\sigma_s}\sum\limits_{i}e^{-\frac{(i\Delta)^2}{2\sigma_s^2}}\mathcal{I}_2\left(\frac{1}{2\sigma_s^2},-\frac{i\Delta}{\sigma_s^2},-\frac{\Delta}{2},\frac{\Delta}{2}\right).
\end{align}
Since solving (\ref{Eq:1}) for equality with respect to $d,~\Delta$ and $\Delta_v$ is cumbersome, we resort to numerical techniques to find the $\Delta$ and $\Delta_v$ parameters that satisfy the average power constraint. The received signal is given by
\begin{align}
Y&= X+S+W \nonumber \\
&=T(V)- (S~\text{mod}~\Delta) +S+W \nonumber \\
&=T(V)+\mathcal{Q}(S)+W. \label{Eq:RX_1DL}
\end{align}

MMSE estimation is directly applied on the received signal to reconstruct the transmitted source sample:
\begin{align}\nonumber
g(y)&=\frac{\int \int vf_{V}(v)f_{S}(s)f_{Y|V,S}(y|v,s)dvds}{\int \int f_{V}(v)f_{S}(s)f_{Y|V,S}(y|v,s)dvds}\\\nonumber
&=\frac{\sum\limits_{i}p(q_i)\int vf_{V}(v)f_{W}(y-T(v)-q_i)dv}{\sum\limits_{i}p(q_i)\int f_{V}(v)f_{W}(y-T(v)-q_i)dv}\\\label{eq:MMSE}
&=\frac{\sum\limits_{i}p(q_i)\cdot\left(\mathbf{F}_{y,q_i,\frac{\kappa}{\Delta_v}}+
\frac{ e^{-\frac{\Delta_v^2}{8}}}{\sqrt{2\pi}}\cdot\left(f_W\left(y-\frac{\kappa}{2}-q_i\right)-f_W\left(y+\frac{\kappa}{2}-q_i\right)\right)
\right)}
{\sum\limits_{i} p(q_i)\cdot\left(\mathbf{G}_{y,q_i,\frac{\kappa}{\Delta_v}}+
\text{Q}(\frac{\Delta_v}{2})\cdot\left(f_W\left(y-\frac{\kappa}{2}-q_i\right)+f_W\left(y+\frac{\kappa}{2}-q_i\right)\right)
\right)},
\end{align}
where $q_i\triangleq i\cdot \Delta$, $i \in \mathbb{Z}$, are the points to which the interference is concentrated when we have $s\in\omega_i,~\omega_i=\left[q_i-\frac{\Delta}{2},q_i+\frac{\Delta}{2}\right)$), and we have
\begin{align}\nonumber
p(q_i)&=\int\limits_{\omega_i}f_S(s)ds= \mathcal{I}_0\left(\frac{\sigma_s^2}{2},q_i,-\frac{\Delta}{2\sigma_s^2},\frac{\Delta}{2\sigma_s^2}\right)\\\label{eq:3}
\quad \mathbf{F}_{y,q_i,\frac{\kappa}{\Delta_v}} &\triangleq \frac{e^{-\frac{(y-q_i)^2}{2\sigma_n^2}}}{2\pi\sigma_n}\cdot\mathcal{I}_1\left(\frac{\Delta_v^2\sigma_n^2+\kappa^2}{2\Delta_v^2\sigma_n^2},\frac{\kappa (y-q_i)}{\Delta_v\sigma_n^2},-\frac{\Delta_v}{2},\frac{\Delta_v}{2}\right),\\\label{eq:5}
\mathbf{G}_{y,q_i,\frac{\kappa}{\Delta_v}} & \triangleq \frac{\Delta_ve^{-\frac{\Delta_v^2(y-q_i)^2}{2\left(\Delta_v^2\sigma_n^2+\kappa^2\right)}}}{\sqrt{2\pi\left(\Delta_v^2\sigma_n^2+\kappa^2\right) }}\cdot\mathcal{I}_0\left(\frac{\Delta_v^2\sigma_n^2+\kappa^2}{2\Delta_v^2\sigma_n^2},\frac{\kappa (y-q_i)}{\Delta_v\sigma_n^2},-\frac{\Delta_v}{2},\frac{\Delta_v}{2}\right).
\end{align}

Finally, the corresponding average distortion is evaluated as below
\begin{align}\label{Eq:Distortion_measure}\nonumber
D&\stackrel{\mathclap{\mbox{($a$)}}}=1-\mathbb{E}[V\hat{V}]\\\nonumber
&=1-\sum\limits_{i}p(q_i)\int \int v g(T(v)+q_i+w) f_W(w)f_V(v)dwdv\\\nonumber
&=1-\sum\limits_{i}p(q_i)\cdot\left({ e^{-\frac{\Delta_v^2}{8}}\cdot\int \left(g\left(\frac{\kappa}{2}+w+q_i\right)-g\left(-\frac{\kappa}{2}+w+q_i\right)\right)f_W(w)dw}\right.\\
&\quad\quad\quad\quad\quad\quad\quad\quad\left.-\int\limits_{-\frac{\Delta_v}{2}}^{\frac{\Delta_v}{2}}\int vf_V(v)g\left(\frac{\kappa}{\Delta_v} v+w+q_i\right)f_W(w)dwdv\right),
\end{align}
where $(a)$ is due to the MMSE estimation.

\begin{rem}
For the ICO scheme (as well as the other non-linear encoding schemes introduced later in this section) it is possible to use an alternative suboptimal decoding scheme called MAP-MMSE. In MAP-MMSE, we first decode the interference concentration index $\hat{q}_i$ using maximum aposteriori (MAP) decoding at the receiver, and then cancel the interference from the received signal $y=T(v)+q_i+w$. Finally MMSE estimation is applied on the remaining signal $y_{\textrm{new}}=T(v)+w+q_i-\hat{q}_i$ to estimate the source sample. This algorithm, in addition to being suboptimal, is also computationally more demanding due to the increased computational complexity in the MMSE stage.  Even considering maximum likelihood (ML) decoder instead of MAP decoder does not improve the computation time significantly. This is because of the dependence of the noise signal with the signal produced by the MAP decoder, $y_{\text{new}}$, which increases the complexity of computing $f_{Y_{\textrm{new}}|V}(y_{\textrm{new}}|v)$. Hence, we restrict our numerical analysis to MMSE estimation as it provides the optimal performance with reduced complexity.
\end{rem}

\subsection{Comparison of ICA and ICO in the Asymptotic Zero-Noise Regime}\label{sec:Comparison}
In order to illustrate the benefits of ICO over ICA we consider the asymptotic zero-noise regime, i.e., we assume that $\sigma_n^2 \rightarrow 0$. For ICA, one can see that,  if $P\geq \sigma_s^2$ then the interference can be completely removed using part of the available power, and zero distortion is achieved in the limit as the noise disappears. On the other hand, when $P< \sigma_s^2$, the best achievable distortion is $(\sigma_s^2-P)/ \sigma_s^2$ (this can be easily verified from (\ref{Eq:b_star}) and (\ref{Eq:D_ICA})); that is, there is always residual distortion in the estimation even if there is no noise in the system.

On the other hand, one can show that in the asymptotic zero-noise regime, independent of the input power constraint, zero distortion can be achieved by the ICO scheme. In the absence of noise, since the received signal is always within the quantization region of the interference signal, the quantization index can always be detected correctly. Once the quantization index is known, the effect of interference can be completely removed.

With MSE estimation applied on the received noiseless signal $T(V)+\mathcal{Q}(S)$, the reconstructed source samples can be written as below
\begin{align}\nonumber
g\left(T(v)+\mathcal{Q}(S)\right)&=\left\{\begin{array}{ccc}
              \frac{e^{-\frac{\Delta_v^2}{8}}}{\sqrt{2\pi}} \quad&,&\quad \text{if}\quad v\geq \frac{\Delta_v}{2},\\
              v \quad&,&\quad\text{if} -\frac{\Delta_v}{2}\leq v < \frac{\Delta_v}{2},\\
              -\frac{e^{-\frac{\Delta_v^2}{8}}}{\sqrt{2\pi}}\quad&,&\quad \text{if}\quad v< -\frac{\Delta_v}{2}.
            \end{array}\right.
\end{align}

The remaining distortion is only due to the companding of the source samples to squeeze them into the quantization region. By letting $\Delta_v$ go to infinity we can reconstruct the source perfectly, and zero distortion can be achieved asymptotically. Note that, letting $\Delta_v \rightarrow \infty$ also means that the average input power depends only on $\Delta$ in the limit.

These arguments show that the ICO scheme can provide significant improvements compared to ICA, particularly when the interference is strong and the noise in the system is low. In the following, we provide other techniques based on the idea of providing a structure to the interference. We will observe that these techniques will further improve the performance of the ICO scheme.

\subsection{One Dimensional Lattice (1DL)}\label{Sec:1DL}

The idea of using a lattice structure for communication in the presence of known interference has been considered in \cite{Erez_shamai_zamir_2005} for the channel coding problem. Here we consider using a similar lattice structure for JSCC. The channel input for the 1DL scheme is given by
\begin{equation}
X=(T(V)-S) ~\text{mod}~\Delta,
\end{equation}
where $T(\cdot)$ is as defined in (\ref{eq:scaling}). In the 1DL scheme, the term $T(v)-s$ is concentrated to one of the quantization points in $\{i\cdot\Delta\}_{i=-\infty}^{\infty}$.

%In order to satisfy the average power constraint, we need to characterize the pdf of $X$. The pdf of $X$ is obtained as follows:
%\begin{eqnarray}\label{1DL_cdf}
%f_{X}(x)=\sum\limits_{i=-\infty}^{\infty}f_{T,S}(i\Delta+x),\quad \text{if}\quad -\frac{\Delta}{2}\leq x<\frac{\Delta}{2},
%\end{eqnarray}
%where $f_{T,S}$ is as follows
%\begin{align}\label{Lattice distribution 1DL}
%%f_{U}(u) &= f_{T}(v)* f_{S}(s) \nonumber \\
%f_{T,S}(u)&=\frac{\Delta_ve^{-\frac{\Delta_v^2u^2}{2\left(\Delta_v^2\sigma_s^2+\kappa^2\right)}}}{\sqrt{2\pi\left(\Delta_v^2\sigma_s^2+\kappa^2\right) }}\mathcal{I}_0\left(\frac{\sigma_s^2\Delta_v^2+\kappa^2 }{2\kappa^2 \sigma_s^2},\frac{u}{\sigma_s^2},-\frac{\kappa}{2},\frac{\kappa}{2}\right)+\frac{Q(\frac{\Delta_v}{2}) }{\sqrt{2\pi}\sigma_s} \left(e^{-\frac{(u-\frac{\kappa}{2})^2} {2\sigma_s^2}}+e^{-\frac{(u+\frac{\kappa}{2})^2}{2\sigma_s^2}}\right).
%\end{align}
In order to satisfy the average power constraint, we need to characterize the pdf of $X$, which can be obtained as follows:
\begin{align}
f_{X}(x)=\left\{\begin{array}{ccl}
           \sum\limits_{i}f_{T\ast S}(i\Delta+x),&\text{if}&\quad -\frac{\Delta}{2}\leq x<\frac{\Delta}{2}, \\
           0,& \text{if}&\quad \text{otherwise}
         \end{array}\right.
\end{align}
where $f_{T\ast S}(u)$ is defined as
\begin{align}
f_{T\ast S}(u)&=\frac{\Delta_ve^{-\frac{\Delta_v^2u^2}{2\left(\Delta_v^2\sigma_s^2+\kappa^2\right)}}}{\sqrt{2\pi\left(\Delta_v^2\sigma_s^2+\kappa^2\right) }}\mathcal{I}_0\left(\frac{\sigma_s^2\Delta_v^2+\kappa^2 }{2\kappa^2 \sigma_s^2},\frac{u}{\sigma_s^2},-\frac{\kappa}{2},\frac{\kappa}{2}\right)+\frac{Q(\frac{\Delta_v}{2}) }{\sqrt{2\pi}\sigma_s} \left(e^{-\frac{(u-\frac{\kappa}{2})^2} {2\sigma_s^2}}+e^{-\frac{(u+\frac{\kappa}{2})^2}{2\sigma_s^2}}\right).
\end{align}

Notice that in 1DL, the channel input $X$ is limited to $[-\Delta/2,\Delta/2)$. The quantization step size, $\Delta$, must be chosen such that the channel input power constraint is satisfied. We recall here that due to the fact that finding a closed form expression for channel power constraint with respect to $\Delta$ and $\Delta_v$ is cumbersome, we resort to numerical calculations to evaluate the value of $\Delta$ that satisfies the power constraint.

The received signal can be written as
\begin{align}
Y&= X+S+W  \nonumber\\
&= (T(V)-S) ~\text{mod}~\Delta+S+W \nonumber \\
&= -\left[ T(V)-S- (T(V)-S) ~\text{mod}~\Delta \right]+T(V)+W  \nonumber \\
&= T(V)-\mathcal{Q}(T(V)-S)+W. \label{eq:RXsignal}
\end{align}

The numerical results for the 1DL scheme are presented in Section \ref{Sec:simulation}. Here we just state that the 1DL scheme achieves lower MMSE compared to ICO, since 1DL supports a larger $\Delta$ value for an equal power constraint. This is mainly due to the bounded channel input which leads to a more efficient use of the available power.

\subsection{ICO with Non-Uniform Quantizer (ICO-NU)}\label{Sec:ICO_NU}

Note that both the ICO and 1DL schemes give some shape to the interference, rather than simply reducing its variance as in ICA. Both schemes use uniform quantization for this. In this section we consider using a non-uniform quantizer for the ICO scheme, and different companders for the source sample depending on the interference signal.

In classical scalar quantization, non-uniform quantization is employed in order to reduce the quantization noise for the more likely values of the underlying signal at the expense of the less likely values. With such a quantizer, in our setting, we would have smaller intervals around zero, and the interval size would increase as we go further away from the origin. Note that, this would reduce the transmission power allocated for interference concentration, since it achieves a lower quantization noise variance. However, this would also mean that we have to compress the source signal even further when the interference realization is close to zero. We observe that the final distortion benefits more from increasing $\Delta$; that is, having quantization points with larger separation. Hence, we apply the opposite of classical non-uniform scalar quantization, and use a lower resolution quantization for more likely values of the interference,  and decrease the quantization interval size as we go further away from zero.

As before, the interference signal is concentrated to the middle point of the quantization interval into which it falls. Since the length of the quantization interval depends on the realization of the interference, a different compander function will be used for each interval. We denote by $\mathcal{Q}^N(\cdot)$ the non-uniform quantizer with decision intervals $\omega_i$ defined as
\begin{align}\label{Eq:events}\nonumber
\omega_0 & \triangleq\{s:-\frac{\Delta_0}{2}\leq s<\frac{\Delta_0}{2}\},\quad i=0,\\
\omega_i & \triangleq\{s:B_i \leq s < B_i +\Delta_i \},\quad i=1,2,...,\\
\omega_i & \triangleq\{s:-B_i-\Delta_i \leq s < -B_i\},\quad i=-1,-2,...,
\end{align}
and quantizations indices $q^N_i$ corresponds to the middle point of each interval. We have
\begin{align}\label{Eq:events}\nonumber
q_{i}^N &= \text{sgn}(i)\cdot\left(B_i+\frac{\Delta_i}{2}\right),
\end{align}
where $\text{sgn}(\cdot)$ is the sign function\footnote{We have $\text{sgn}(x)=1$ if $x>0$, $-1$ if $x<0$, and $0$ if $x=0$.}, and
\begin{align}
B_i \triangleq\left\{\begin{array}{ccl}
        |i|\cdot\Delta_0/2,& \text{if}\quad\quad i=\{1,0,-1\},\\
        \frac{\Delta_0}{2}+\sum\limits_{j=1}^{|i|-1}\Delta_j,& \quad\quad \text{otherwise}.
      \end{array}\right.
\end{align}

We define the function $\bar{f}_S(s)$ as follows
\begin{eqnarray}\label{Eq:criterion1}
\bar{f}_S(s)\triangleq\frac{1}{f_S(s)^a},
\end{eqnarray}
where $f_S(s)$ is the pdf of the interference $S$, and $a\geq0$ is a parameter to be optimized. The length of the i-th quantization interval, $\Delta_i$, is chosen such that
\begin{eqnarray}\label{Eq:criterion}
2\int\limits_{s\in \omega_0}\bar{f}_S(s)ds=\int\limits_{s\in\omega_i}\bar{f}_S(s)ds, \quad  i=1,2,\ldots
\end{eqnarray}

For Gaussian interference and $a\geq 0$ it can be shown that
\begin{align}
|i|>|j| &\Rightarrow \Delta_{i}\leq\Delta_{j},\\
|i|=|j| &\Rightarrow \Delta_{i}=\Delta_{j}.
\end{align}

%----------------------------------------------
%\begin{figure*}[!t]
%\begin{equation}\label{Eq:f_X_1DL}
%f_{X_{1DL}}(x) = \sum\limits_{i=-\infty}^{\infty}\frac{e^{-\frac{(i\Delta+x)^2}{2(\sigma_s^2+\alpha^2\sigma_v^2)}}}{2\sqrt{2\pi (\sigma_s^2+\alpha^2\sigma_v^2)}}\left(\text{erf}(\frac{\frac{\Delta}{2}-m_{l_2}}{\sqrt{2}\sigma_{l_2}})-\text{erf}(\frac{-\frac{\Delta}{2}-m_{l_2}}{\sqrt{2}\sigma_{l_2}})\right)
%+\frac{Q_{\sigma_v^2}(\frac{\Delta_v}{2}) }{\sqrt{2\pi}\sigma_s}\left(e^{-\frac{(i\Delta+x-\frac{\Delta}{2})^2}{2\sigma_s^2}}+e^{-\frac{(i\Delta+x+\frac{\Delta}{2})^2}{2\sigma_s^2}}\right).
%\end{equation}
%\hrule
%\end{figure*}
%----------------------------------------------

At the transmitter, if $S$ falls into the quantization interval $\omega_i$, we have $\mathcal{Q}^N(S)=q_{i}^N $. Therefore, source $V$ is transformed as follows
\begin{align}
\label{eq:scaling_nonuniform}
T(v, q_{i}^N)&\triangleq\left\{ \begin{array}{ccc}
           \frac{\Delta_i}{2}, &\text{if}&\quad v\geq \frac{\Delta_v}{2} \\
           \frac{\Delta_i}{\Delta_v} v, &\text{if}&\quad -\frac{\Delta_v}{2}\leq v<\frac{\Delta_v}{2}\\
           -\frac{\Delta_i}{2}, &\text{if}&\quad v< -\frac{\Delta_v}{2}
         \end{array}\right.,
\end{align}
where we have defined $T(v, q_{i}^N)$ to denote the companding function, in order to highlight its dependence on the realization of the interference quantization $q_i^N$. The transmitted signal is generated as below
\begin{eqnarray}
X=T(V, \mathcal{Q}^N(S))-(S~\text{mod}~\Delta^N),
\end{eqnarray}
where $(S~\text{mod}~\Delta^N)$ denotes the quantization noise for the non-uniform scalar quantizer with $\mathcal{Q}^N(\cdot)$.

To satisfy the average power constraint we follow the same approach as in Section \ref{Sec:ICO}. For brevity we define $U\triangleq (S~\text{mod}~\Delta^N)$. We have
\begin{align}\nonumber
\mathbb{E}[X^2]&=\sum_{i}p(q_{i}^N)\cdot\int_{-\Delta_i}^{\Delta_i}x^2f_{X|\mathcal{Q}^N}\left(x|\mathcal{Q}^N(S)=q_{i}^N\right)dx\\\nonumber
&=\sum_{i}p(q_{i}^N)\cdot\left(\sigma_{T(V,q_{i}^N)}^2+\sigma_{U|\mathcal{Q}^N(S)=q_{i}^N}^2\right),\\\label{Eq:Non uniform APC}
&=\sum_{i}p(q_{i}^N)\cdot\sigma_{T(V,q_{i}^N)}^2+\sigma_{U}^2,
\end{align}
where
\begin{align}\nonumber
p(q_{i}^N) &=\mathcal{I}_0\left(\frac{\sigma_s^2}{2},q_{i}^N,-\frac{\Delta_i}{2\sigma_s^2},\frac{\Delta_i}{2\sigma_s^2}\right)\\
\sigma_{T(V,q_{i}^N)}^2&=\Delta_i^2 \left[\frac{1}{\Delta_v^2}+Q\left(\frac{\Delta_v}{2}\right)\left(\frac{1}{2}-\frac{2}{\Delta_v^2}\right)-\frac{e^{-\frac{\Delta_v^2}{8}}}{\Delta_v\sqrt{2\pi}}\right].
\end{align}

To evaluate $\sigma_{U}^2$ in (\ref{Eq:Non uniform APC}), we need the distribution of $U$ for the non-uniform quantizer. Since $\max\limits_{i}\{\Delta_i\}=\Delta_0$, we have $U\in[-\Delta_0,\Delta_0)$. The \emph{cumulative distribution function} (cdf) of $U$ can be written as
\begin{eqnarray}\label{Eq:13}
F_{U}(u)=\sum_{i}F_{U|\mathcal{Q}^N}\left(u|\mathcal{Q}^N(S)=q_{i}^N\right)p(q_{i}^N),
\end{eqnarray}
where $F_{U|\mathcal{Q}^N}\left(u|\mathcal{Q}^N=q_{i}^N\right)$ for different $i$'s can be expanded as below
\begin{align}\nonumber
F_{U|\mathcal{Q}^N}\left(u|\mathcal{Q}^N(S)=q_{i}^N\right)&=\frac{1}{p(q_{i}^N)}\int_{q_{i}^N-\frac{\Delta_i}{2}}^{q_{i}^N+u}f_S(s)ds,\quad -\frac{\Delta_i}{2}\leq u < \frac{\Delta_i}{2}.\nonumber
%F_{U_2|\mathcal{Q}^N}(u|q_{N_i})&=&\frac{1}{p(q_i)}\left[\int_{-A_i-\frac{\Delta_i}{2}}^{-A_i+u}f_S(s)ds+\int_{A_i-\frac{\Delta_i}{2}}^{A_i+u}f_S(s)ds\right]\\\nonumber
%&=&\frac{1}{p(q_i)}\int_{A_i-\frac{\Delta_i}{2}}^{A_i+u}f_S(s)+f_S(s-2A_i)ds,\quad -\frac{\Delta_i}{2}\leq u < \frac{\Delta_i}{2},~i=2,3,...,\label{eq:tale1}
\end{align}
%where, in the above equations for $i\geq2$, we have $A_i\triangleq B_i+\Delta_i/2$ .

By differentiating (\ref{Eq:13}) with respect to $u$, and recalling that $q_{-i}^N=-q_{i}^N$ and $p(q_{-j}^N)=p(q_{j}^N)$, we obtain
\begin{align}\nonumber
\Scale[1.1]{f_{U}(u)=\frac{f_S(u)}{p(q_{0}^N)}+\sum\limits_{i=1}^{\infty}\frac{f_S(-q_{i}^N+u)\cdot R\left(\frac{2u+B_i+B_{i+1}}{2(B_{i+1}-B_i)}\right)+f_S(q_{i}^N+u)\cdot R\left(\frac{2u-B_i-B_{i+1}}{2(B_{i+1}-B_i)}\right)}{p(q_{i}^N)},\quad -\frac{\Delta_0}{2}\leq u < \frac{\Delta_0}{2}.}
\end{align}
Using conventional tools in probability theory, $\sigma_{U}^2$ in (\ref{Eq:Non uniform APC}) can be evaluated as
\begin{align}\nonumber
\sigma_{U}^2&=\sum\limits_{i}\frac{e^{-\frac{\left(q_{i}^{N}\right)^2}{2\sigma_s^2}}}{\sqrt{2\pi}\sigma_s}\mathcal{I}_2(\frac{1}{2\sigma_s^2},\frac{q_{i}^N}{\sigma_s^2},-\frac{\Delta_i}{2},\frac{\Delta_i}{2}).
\end{align}

The received signal for the ICO-NU scheme is given by
\begin{align}\nonumber
Y&=X+S+W\\\nonumber
&=T(V, \mathcal{Q}^N(S))- \left(S~\text{mod}~\Delta^N\right) +S+W\\
&=T(V, \mathcal{Q}^N(S)) + \mathcal{Q}^N(S) + W.
\end{align}

At the receiver we use MMSE estimation as introduced in Section \ref{Sec:ICO}. The source reconstruction and final distortion is obtained as follows.
\begin{align}\label{Eq:V_hat_NUN}
g^{N}(y)&= \frac{\sum\limits_{i}p(q_{i}^N)\cdot\left(\mathbf{F}_{y,q_{i}^N,\frac{\Delta_j}{\Delta_v}}+
\frac{ e^{-\frac{\Delta_v^2}{8}}}{\sqrt{2\pi}}\cdot\left(f_W(y-\frac{\Delta_i}{2}-q_{i}^N)-f_W(y+\frac{\Delta_i}{2}-q_{i}^N)\right)
\right)}
{\sum\limits_{i} p(q_{i}^N)\cdot\left(\mathbf{G}_{y,q_{i}^N,\frac{\Delta_j}{\Delta_v}}+\text{Q}
(\frac{\Delta_v}{2})\cdot\left(f_W(y-\frac{\Delta_i}{2}-q_{i}^N)+f_W(y+\frac{\Delta_i}{2}-q_{i}^N)\right)
\right)},
\end{align}
\begin{align}\label{Eq:Distortion_measure_NUN}
D&=1-\sum\limits_{i}p(q_{i}^N) \int \int vg_{\textrm{mmse}}^{N}(T(v,q_{i}^N)+q_{i}^N+w) f_W(w)f_V(v)dwdv.
\end{align}

\subsection{1DL with Non-Uniform Quantizer (1DL-NU)}\label{sec:1DL-NU}

In this section we consider the 1DL scheme combined with a non-uniform quantizer similarly to the ICO scheme in Section \ref{Sec:ICO_NU}. The transmitted signal is given as below
\begin{eqnarray}
X_{\text{1DL}}=\left([T(V,\mathcal{Q}^N(S))-S]~\text{mod}~\Delta^{N}\right).
\end{eqnarray}

To satisfy the average power constraint we follow the same approach as in Section \ref{Sec:ICO}. We have
\begin{align}\nonumber
\mathbb{E}[X^2]&=\sum_{i}p(q_{i}^N)\cdot\int_{-\frac{\Delta_i}{2}}^{\frac{\Delta_i}{2}}x^2f_{X|\mathcal{Q}^N}\left(x|\mathcal{Q}^N(S)=q_{i}^N\right)dx.
\end{align}
where
\begin{align}\nonumber
f_{X|\mathcal{Q}^N}\left(x|\mathcal{Q}^N(S)=q_{i}^N\right)&=f_{c}^{q_{i}^N}(q_{i}^N+x)+f_{c}^{q_{i}^N}(q_{i}^N+x-\text{sgn}(x)\Delta_i),\quad -\frac{\Delta_i}{2}\leq x < \frac{\Delta_i}{2}\quad \text{for all }i,
\end{align}
where $f_{c}^{q_{i}^N}(u)=f_{T(V,q_{i}^N)}(v)\ast f_{S|\mathcal{Q}^N}(s|\mathcal{Q}^N(S)=q_{i}^N)$, and we have
\begin{align}\nonumber
f_{c}^{q_{i}^N}(u)&=\left\{\begin{array}{ccc}
\alpha_i\mathcal{I}_0\left(\beta_i,\frac{u}{\sigma_s^2},-\frac{\Delta_i}{2},u+\frac{\Delta_i}{2}\right)+\frac{Q\left(\frac{\Delta_v}{2}\right)e^{-\frac{(u+\Delta_i/2)^2}{2\sigma_s^2}} }{p(q_{i}^N)\sqrt{2\pi}\sigma_s} ,& -\Delta_i\leq u <0 \\
\alpha_i\mathcal{I}_0\left(\beta_i,\frac{u}{\sigma_s^2},u-\frac{\Delta_i}{2},\frac{\Delta_i}{2}\right)+
\frac{Q\left(\frac{\Delta_v}{2}\right)e^{-\frac{(u-\Delta_i/2)^2}{2\sigma_s^2}} }{p(q_{i}^N)\sqrt{2\pi}\sigma_s} ,& 0 \leq u<\Delta_i
\end{array}
\right. ,\nonumber
\end{align}
where $\alpha_i=\frac{\Delta_v e^{-\frac{\Delta_v^2u^2}{2(\Delta_v^2\sigma_s^2+\Delta_i^2)}}}{p(q_{i}^N)\sqrt{2\pi(\Delta_v^2\sigma_s^2+\Delta_i^2)}}$, and $\beta_i=\frac{\Delta_v^2\sigma_s^2+\Delta_i^2}{2\sigma_s^2\Delta_i^2}$

The received signal for the 1DL-NU scheme is given by
\begin{align}\nonumber
Y&=X+S+W\\\nonumber
&=\left([T(V,\mathcal{Q}^N(S))-S]~\text{mod}~\Delta^{N}\right)+S+W\\
&=T(V,\mathcal{Q}^N(S)) - \mathcal{Q}^N(S) + W.
\end{align}

At the receiver we use MMSE estimation as introduced in Section \ref{Sec:ICO}. To reconstruct the source samples at the receiver we use (\ref{Eq:V_hat_NUN}) and (\ref{Eq:Distortion_measure_NUN}).

\begin{rem}
The intuition behind choosing the function in \ref{Eq:criterion1} is the following. We know that clipping the source sample injects a distortion at the encoder side. Therefore, given an average power constraint, the larger the $\Delta_v$ value, the smaller the distortion introduced by clipping. Since the interference has a Gaussian distribution, we know that realizations of the interference around the origin are more likely than those towards the tails of the distribution. Note that the quantization noise variance of the quantization scheme we use, corresponds to the power spent for concentrating the interference at the transmitter. Therefore,  Uniformly quantizing the interference $S$, is equivalent to assigning the power budget uniformly across realizations of the interference, whereas, assigning larger intervals around the origin distributes the power budget non-uniformly among interference realizations, such that, more likely interference realizations are quantized with larger quantization intervals; and hence, they require more power, but allow smaller distortion due to clipping. The heuristic interference quantization scheme used here can be further improved by devising a numerical technique similar to the classical Lloyd-Max algorithm \cite{Lloyd}. We leave the optimization of the interference quantizer for the ICO and 1DL schemes as a future work.
\end{rem}

\section{Necessary Condition for Optimality and Numerically Optimized Encoder (NOE) Design}\label{Sec:Numerically Optimized encoder}
As stated in the Introduction, the optimal zero or low-delay joint source-channel coding scheme is an open problem in most communication scenarios, with few exceptions \cite{Goblick:IT:65}. A common approach for these problems in the literature \cite{Akyol-Viswanatha-Rose-Ramstad}, \cite{Floor-Ramstad-Wernersson} is to formulate the optimal encoder mapping as an unconstrained optimization problem through the Lagrangian, then to apply calculus of variations techniques to obtain a necessary condition for the optimal mapping, and finally numerically obtain an encoder mapping, typically using an iterative steepest descent algorithm, that satisfies this necessary condition. Due to lack of convexity, this solution does not guarantee global optimality, and the final solution is highly sensitive to the initial mapping. Despite these drawbacks, with carefully chosen initial mappings, and sufficiently high-grained quantization of the continuous source and channel alphabets, these numerically optimized encoders (NOEs) achieve the best known performance in most scenarios.

% In this section, we will follow the same approach as in [ABOU SALEH].

In this section, we will follow the same approach as in \cite{Akyol-Viswanatha-Rose-Ramstad}, \cite{saleh_alajaji_2014}. We briefly include the derivations for completeness, and then, we numerically obtain the encoder that satisfies this condition. Numerical techniques have been previously used for joint source-channel mappings in various scenarios in \cite{Akyol-Viswanatha-Rose-Ramstad,Mustafa_Mehmetoglu,Karlsson-Skoglund,Floor-Ramstad-Wernersson}. By writing the Lagrangian cost function for this system model we have
\begin{align}\nonumber
J(h,g)&=\mathbb{E}[(V-\hat{V})^2]+\lambda\cdot \mathbb{E}[h(V,S)^2]\\\label{Eq:2}
&=1-\int \int \left(\int vg(h(v,s)+w+s)f_W(w)dw - \lambda h(v,s)^2\right)f_V(v)f_S(s) dv ds,
\end{align}
where $\lambda$ is the Lagrangian multiplier, and $h(V,S)$ is the encoder mapping function. By writing Euler-Lagrange equations \cite[Section 7.5]{Luenberger_1969} we have
\begin{eqnarray}\label{Eq:4}
\nabla_h J(h,g)=\left(2\lambda h(s,v)-\int vg^{'}(h(v,s)+w+s)f_W(w)dw \right)\cdot f_V(v)f_S(s),
\end{eqnarray}
where $g^{'}(\cdot)$ is the derivative of $g(\cdot)$. From calculus of variations \cite{Luenberger_1969}, it is well-known that for the optimal encoder mapping, (\ref{Eq:4}) must be zero. This yields the necessary condition for optimality as below
\begin{eqnarray}\label{Eq:8}
h(v,s)=\frac{v}{2\lambda}\int g^{'}(h(v,s)+w+s)f_W(w)dw.
\end{eqnarray}

The optimal decoder is the MMSE estimator, and the corresponding distortion is obtained as below
\begin{eqnarray}\nonumber
D=1-\int \int \int  vg(h(v,s)+s+w)f_S(s)f_V(v)f_W(w)dv ds dw.
\end{eqnarray}

\begin{rem}We note here that the uncoded transmission satisfies the necessary condition in (\ref{Eq:8}). Considering the transmitted signal as $h(v,s)=av$, where $a=\sqrt{P}$, the MMSE estimation can be simplified to $g(y)=cy$, where $c=\frac{a}{a^2+\sigma_s^2+\sigma_n^2}$ and $y$ is the received signal. Substituting $h(\cdot)$, $g(\cdot)$ in (\ref{Eq:8}) we have
\begin{eqnarray}
av=\frac{cv}{2\lambda}\int f_W(w)dw~\Rightarrow~\lambda=\frac{c}{2a}.
\end{eqnarray}

Substituting $\lambda=\frac{c}{2a}$ in (\ref{Eq:4}) makes the gradient $\nabla_h J(h,g)$ zero. The final resulting distortion in this case is $D=\frac{1}{1+\frac{P}{\sigma_s^2+\sigma_n^2}}$. It is also noticed that for the ICA scheme, since $\lambda=\frac{a}{c}+\frac{bs}{c}$ is not a constant, that does not satisfy (\ref{Eq:8}).
\end{rem}

\subsection{Numerically Optimized Encoder (NOE)}
Since the optimality condition for the encoder derived above does not have a closed form expression, we use the iterative steepest decent algorithm to obtain the encoder numerically. During the iterations the encoder is updated as below
\begin{eqnarray}\label{Eq:6}
h_{i+1}(v,s)=h_i(v,s)-\mu \nabla_h(h,g),
\end{eqnarray}
where $i$ is the iteration index, $\mu$ is the step size, and $\nabla_h(h,g)$ is obtained as in (\ref{Eq:4}). At each iteration the initial cost (\ref{Eq:2}) is decreasing. Iterations are performed until $\nabla_h(h,g)$ reaches a predefined threshold value.
In order to calculate the integrals in (\ref{Eq:4}) at each iteration we use discretization. It is worth mentioning that, since discretization injects some residual error into the algorithm, it is essential to increase the accuracy in order to make the residual distortion (due to discretization) negligible compared to the final achievable distortion. Hence, the simulation takes considerably longer time to converge at high SNR values.

In our simulations we start from the low power constraints. The algorithm is initiated with a vector whose elements are the different values assigned to each discretized pair of $(v,s)$. For the lowest power constraint, the initial values are chosen close to zero to make sure that they satisfy the average power constraint. The final solution obtained for a low power constraint is used as the initial guess for the higher power constraint, and so on. It should be remarked, that there is no guarantee that this iterative optimization scheme converges to the global optimal solution. We have also tried to initiate the NOE from the encoder mappings obtained for ICO and 1-DL schemes, as well as their non-uniform quantized counterparts; but in all cases we obtained the exact same final encoder mapping.

%----------------------------------------------
\begin{figure}
\begin{centering}
\includegraphics[scale=0.3]{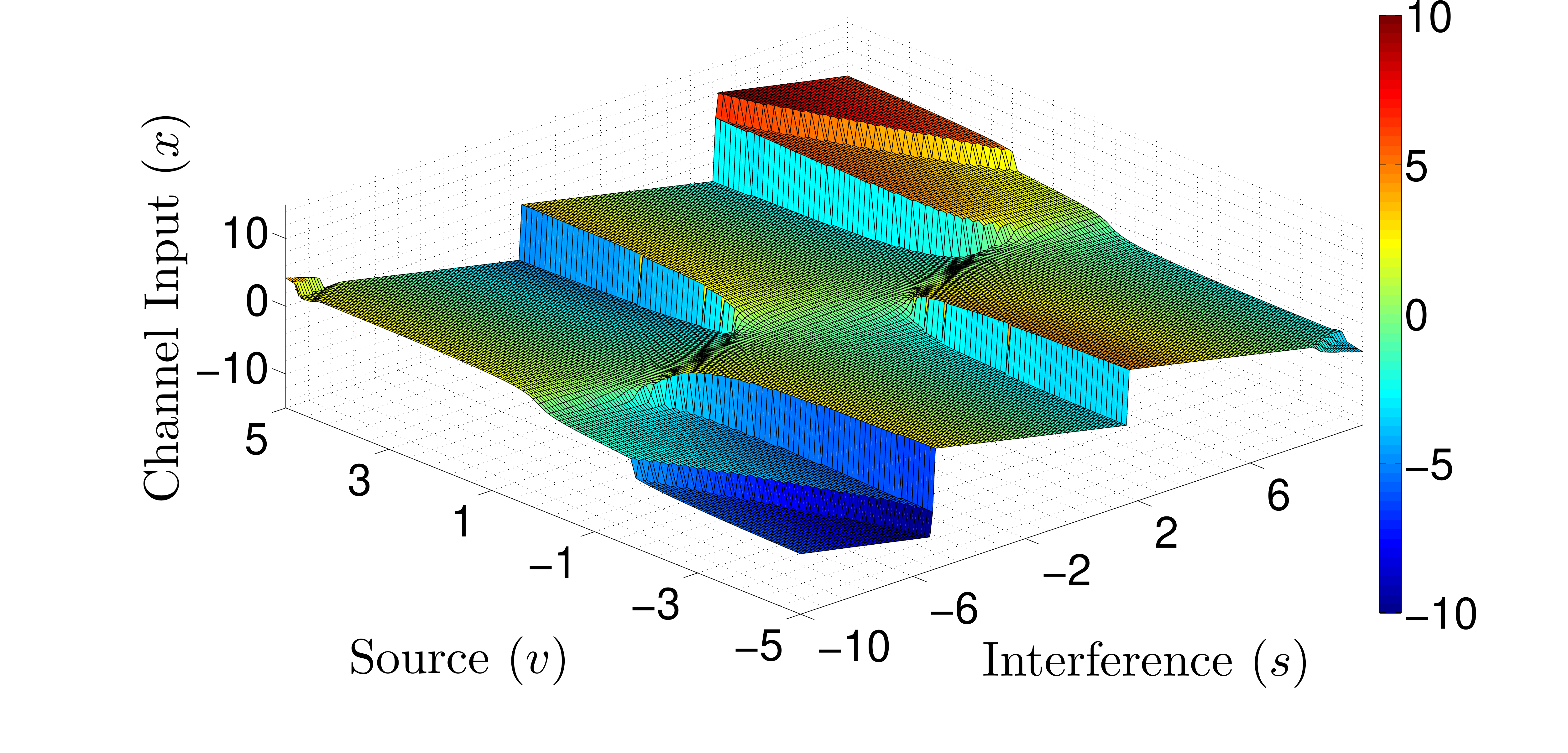}
\par\end{centering}
\caption{\label{Fig_figure7} Encoder mapping h(v,s) for NOE ($\sigma_s^2=4$, $\sigma_n^2=1$).}
\vspace{0mm}
\end{figure}
%----------------------------------------------

%%----------------------------------------------
%\begin{figure}
%\begin{centering}
%\includegraphics[scale=0.4]{figure/received.pdf}
%\par\end{centering}
%\caption{\label{Fig_figure8} Received signal in the absence of the noise $Y=X+S$, for NOE ($\sigma_s^2=4$, $\sigma_n^2=1$).}
%\vspace{0mm}
%\end{figure}
%%-------------

%----------------------------------------------
\begin{figure}
\begin{centering}
\includegraphics[scale=0.5]{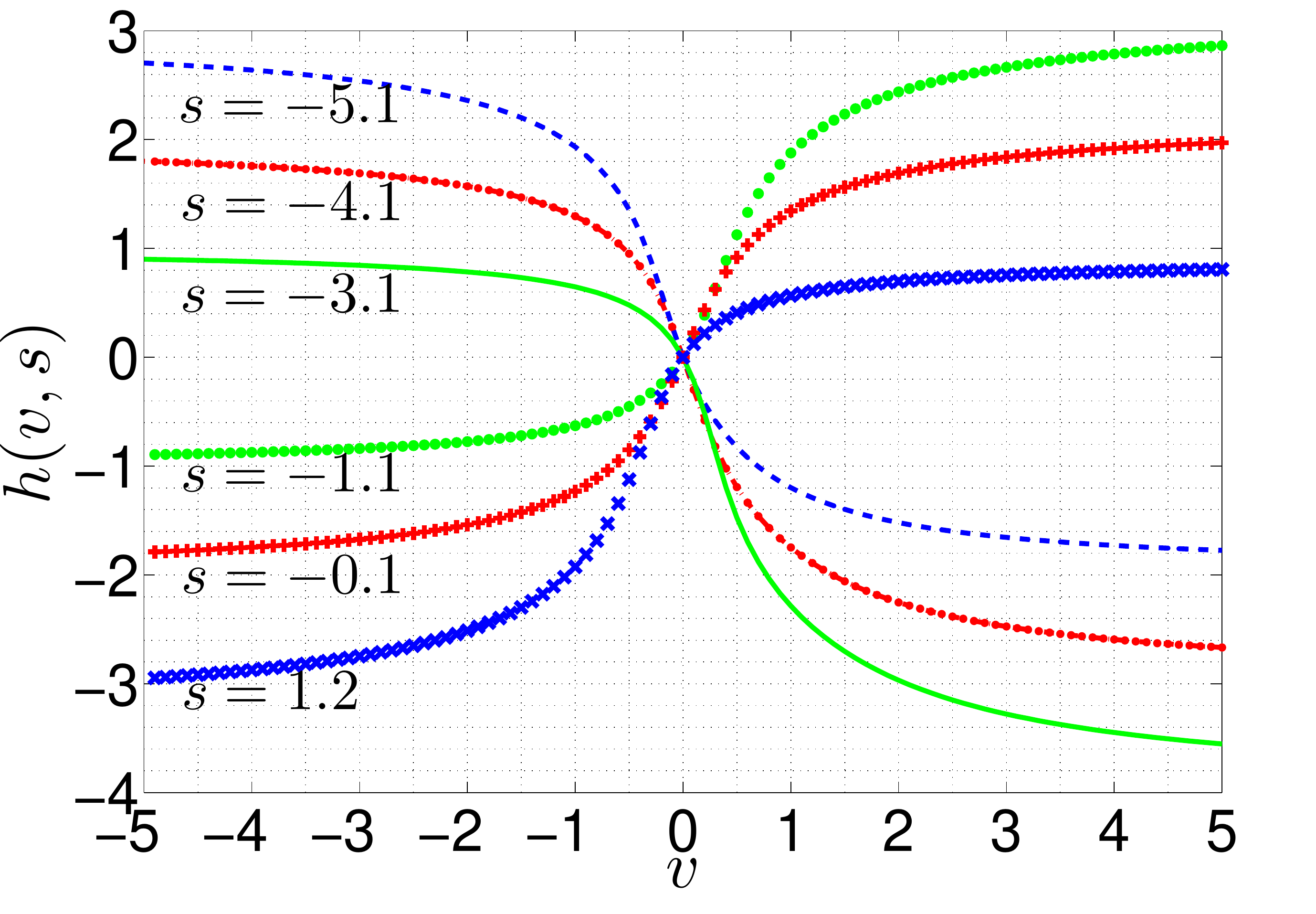}
\par\end{centering}
\caption{\label{Fig_figure21} NOE mapping for different interference values ($\sigma_s^2=4$, $\sigma_n^2=1$).}
\vspace{0mm}
\end{figure}
%-------------

%----------------------------------------------
\begin{figure}
\begin{centering}
\includegraphics[scale=0.5]{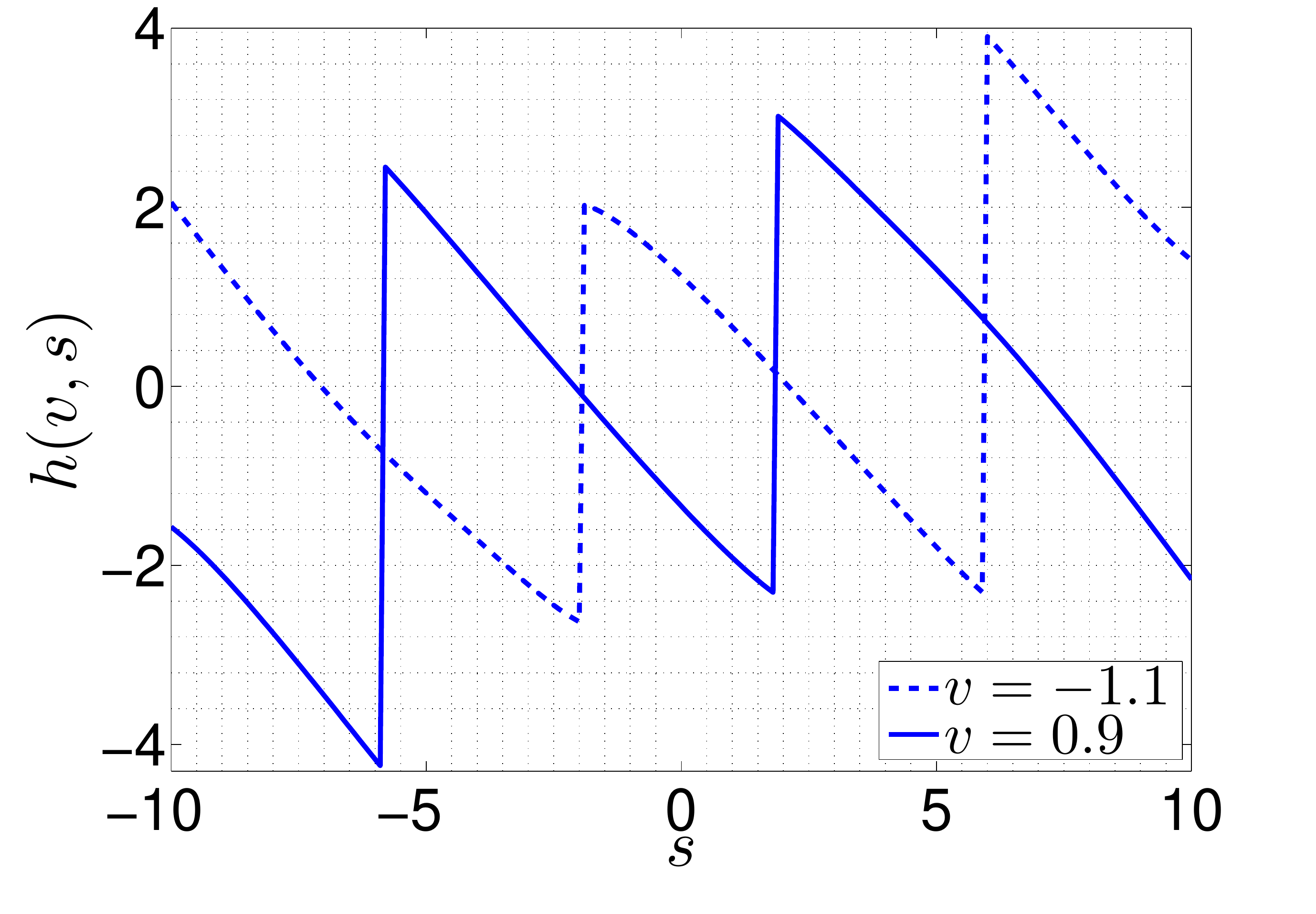}
\par\end{centering}
\caption{\label{Fig_figure20} NOE mapping for source outputs $v=0.9$ and $v=-1.1$ ($\sigma_s^2=4$, $\sigma_n^2=1$).}
\vspace{0mm}
\end{figure}
%-------------

In Fig. \ref{Fig_figure7}, the encoder structure for numerically optimized encoder with $P=4$ dB is shown. The plot shows, in a colour-coded fashion, the channel input value (here in range $[-40,40]$) corresponding to each discretized pair of source $v$ and interference $s$ values. To elaborate the details of Fig \ref{Fig_figure7}, the encoder mapping for different values of the source and the interference outputs are shown   in Fig. \ref{Fig_figure21} and \ref{Fig_figure20}, respectively.
From Figures \ref{Fig_figure7}-\ref{Fig_figure20},
%\ref{Fig_figure7}, Fig. \ref{Fig_figure21} and Fig. \ref{Fig_figure20}
we observe that, \textit{i}) the source is clipped similarly to the parameterized nonlinear schemes considered in Section \ref{Sec:achievable transmission schemes} (see Fig. \ref{Fig_figure21} and Fig. \ref{Fig_figure9} for comparison); \textit{ii}) Depending on which interval the interference falls into, the transmitted signal resembles a shifted version of a linear mapping (see Fig. \ref{Fig_figure20}). This is similar to the transmission of quantization noise in ICO and 1DL schemes.
%(notice the slope of different sections of the mapping in Fig. \ref{Fig_figure7}, which gets flat after being added with interference in Fig. \ref{Fig_figure8}).

\section{Numerical results}\label{Sec:simulation}

%----------------------------------------------
\begin{figure}
\begin{centering}
\includegraphics[scale=0.5]{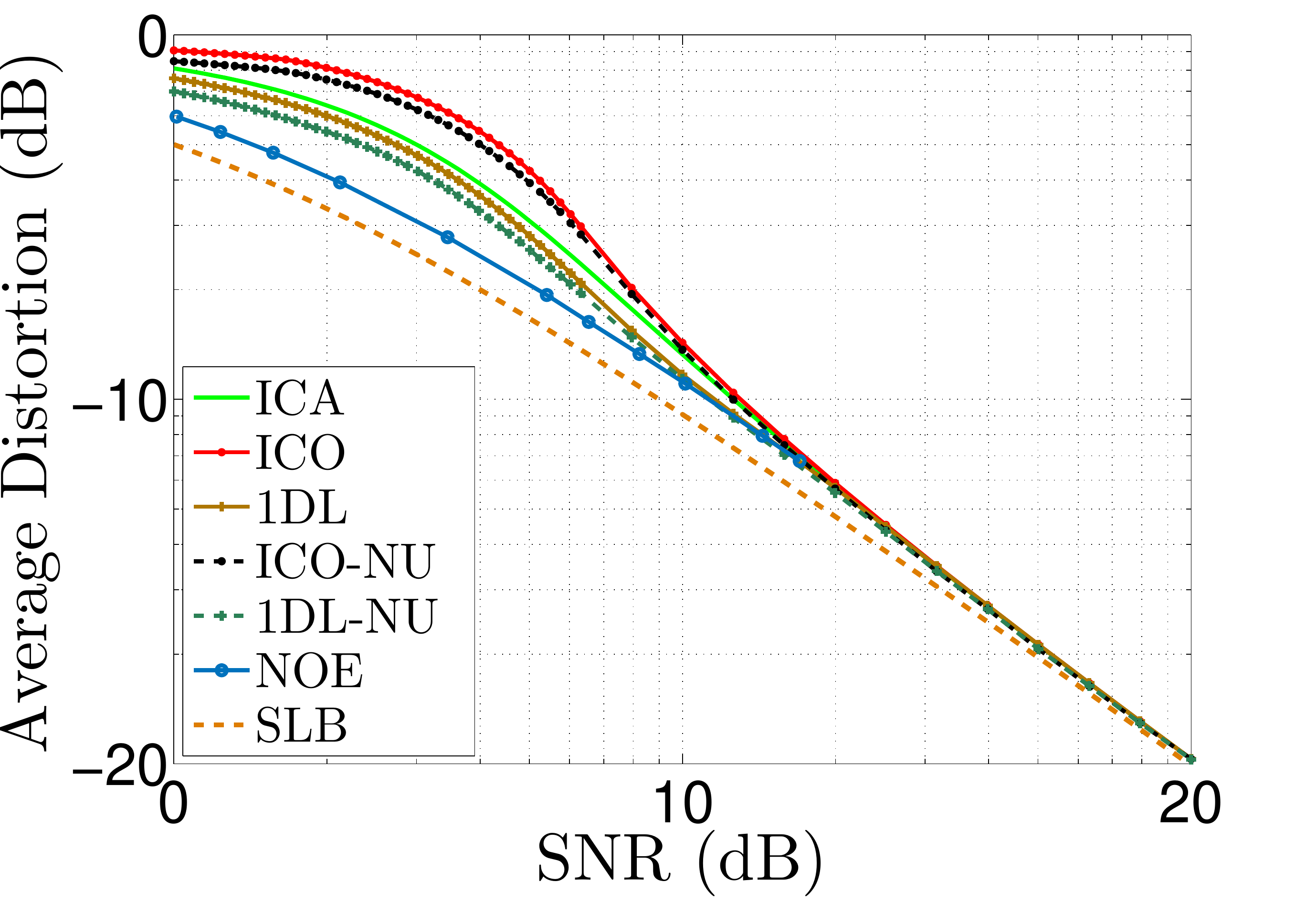}
\par\end{centering}
\caption{\label{Fig_figure1} Average MSE distortion (dB) vs. the average SNR (dB) for the proposed schemes for $\sigma_s^2=4$, $\sigma_n^2=1$ .}
\vspace{0mm}
\end{figure}
%----------------------------------------------

%----------------------------------------------
\begin{figure}
\begin{centering}
\includegraphics[scale=0.5]{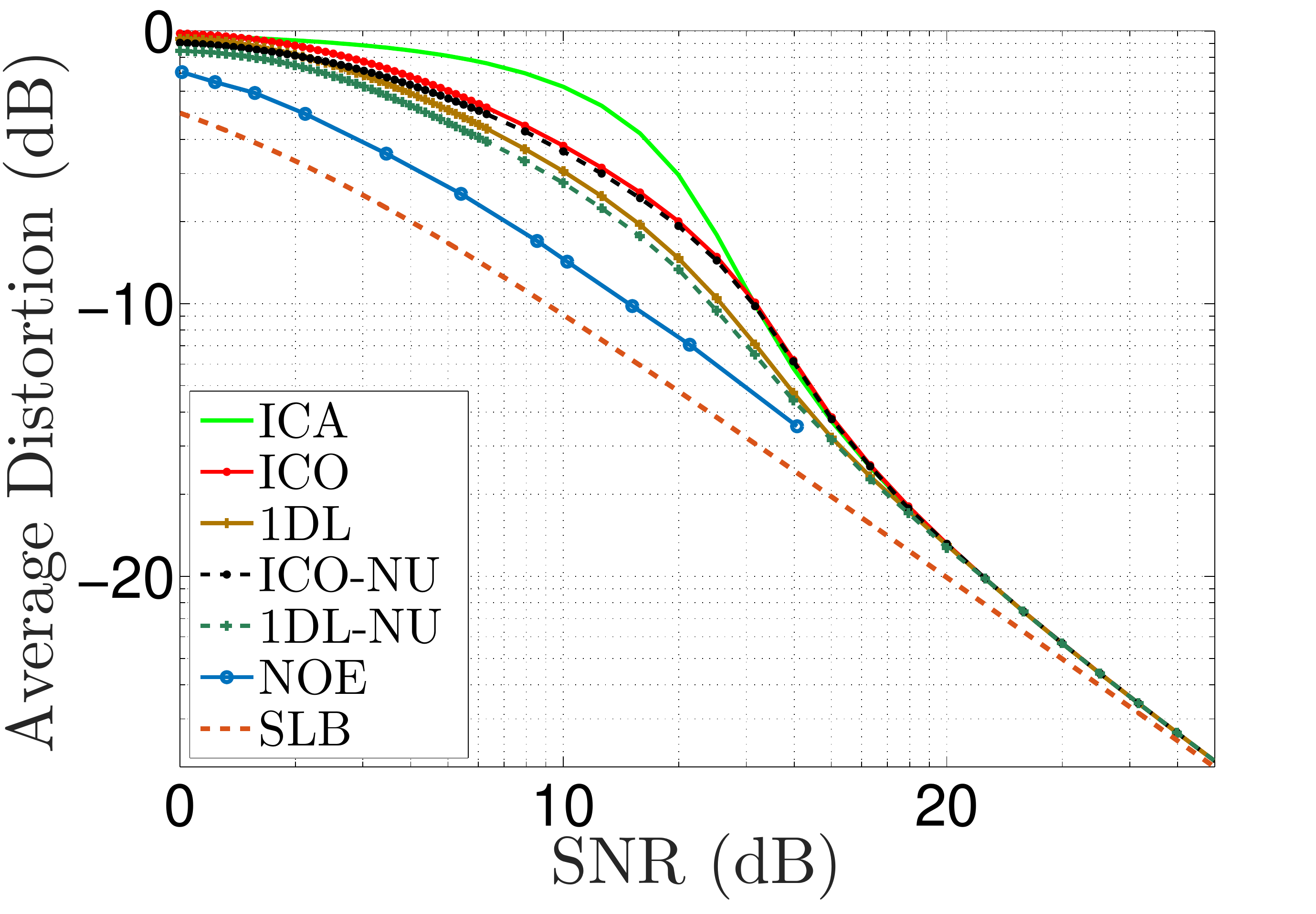}
\par\end{centering}
\caption{\label{Fig_figure2} Average MSE distortion (dB) vs. the average SNR (dB) for the proposed schemes for $\sigma_s^2=25$, $\sigma_n^2=1$ .}
\vspace{0mm}
\end{figure}
%----------------------------------------------

We remark here that obtaining closed-form expressions for the optimal performance of JSCC under strict delay constraints is extremely difficult if not impossible. Instead, in this section, we provide numerical results comparing the performances of the proposed transmission schemes. We will also include the Shannon theoretic lower bound (SLB) obtained by evaluating the rate-distortion function of the Gaussian source at the capacity of the underlying channel when the interference is completely removed. Not surprisingly this lower bound is quite loose in general.
%----------------------------------------------
\begin{figure}
\begin{centering}
\includegraphics[scale=0.5]{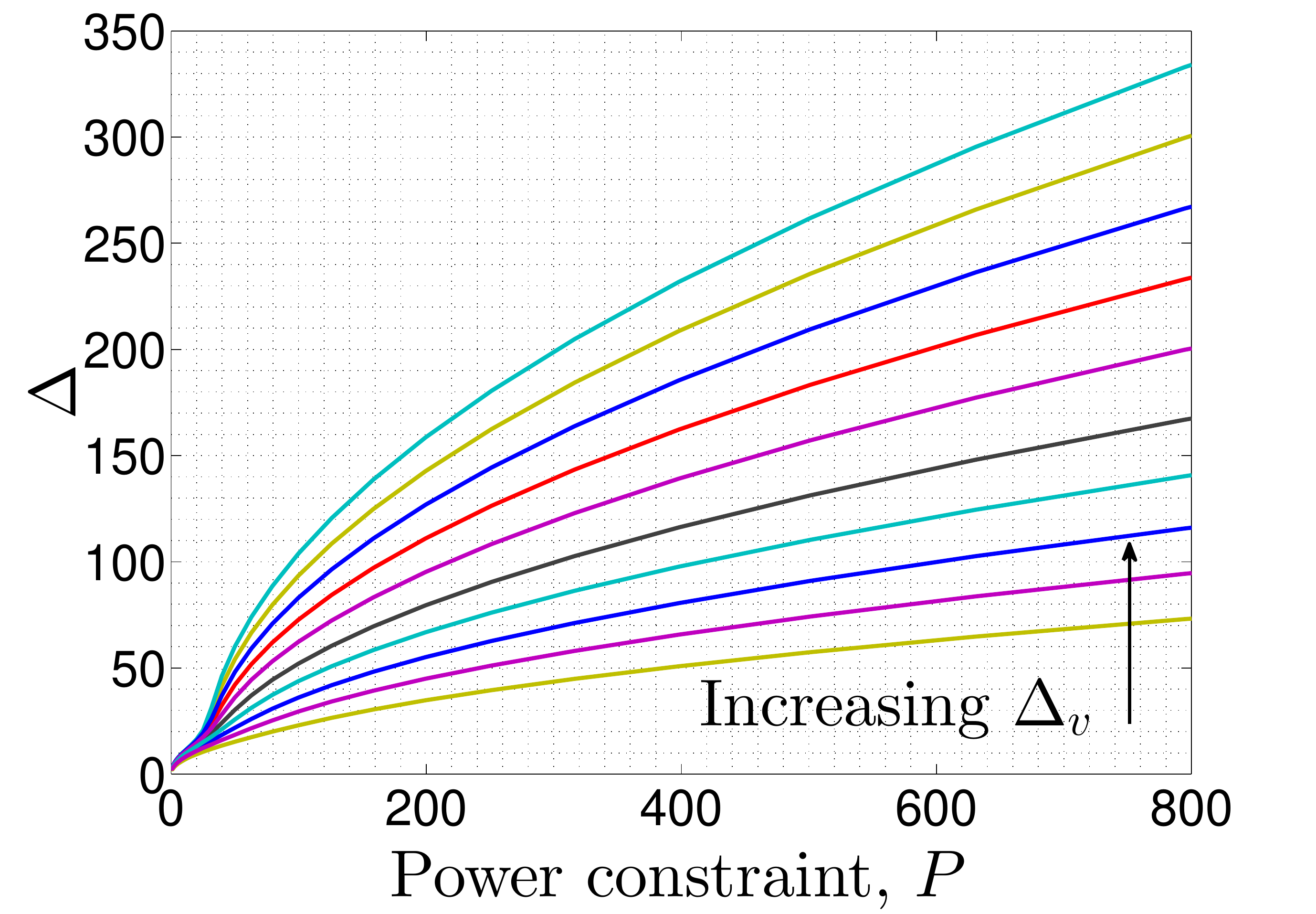}
\par\end{centering}
\caption{\label{Fig_figure3} $\Delta$ vs. power constraint for different values of $\Delta_v$ for ICO ($\sigma_s^2=25$, $\sigma_n^2=1$) .}
\vspace{0mm}
\end{figure}
%----------------------------------------------

%----------------------------------------------
\begin{figure}
\begin{centering}
\includegraphics[scale=0.5]{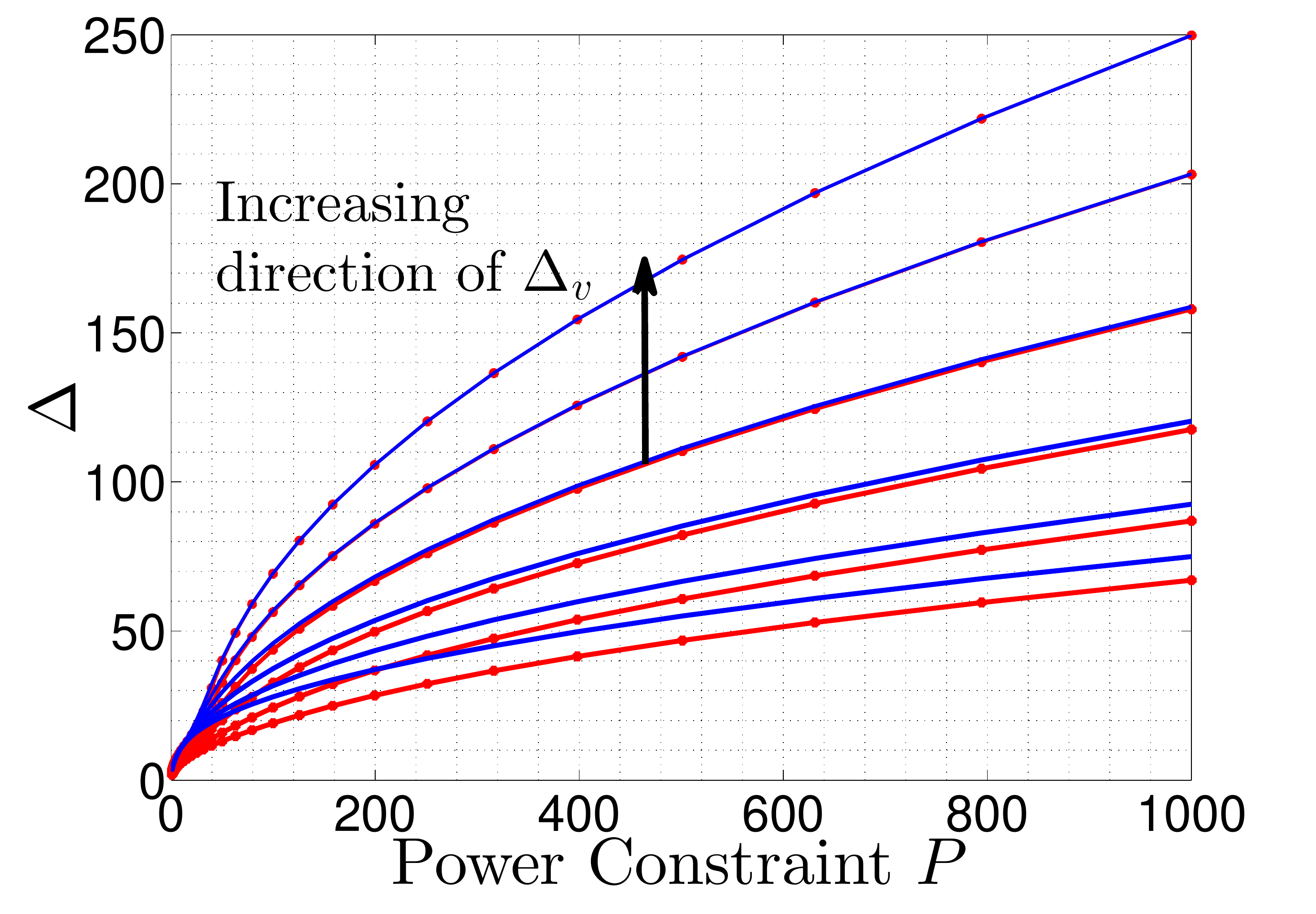}
\par\end{centering}
\caption{\label{Fig_figure4} $\Delta$ vs. channel power constraint for ICO and 1DL for different values of $\Delta_v$ ($\sigma_s^2=25$, $\sigma_n^2=1$). Red dotted lines are associated with ICO and blue lines are associated with 1DL. $\Delta$ for 1DL is always greater than that for ICO (for fixed value of power constraint and $\Delta_v$)  }
\vspace{0mm}
\end{figure}
%----------------------------------------------
In Fig. \ref{Fig_figure1} and \ref{Fig_figure2}, performances of the proposed transmission schemes are illustrated and compared with SLB for different SNR levels. For ICO-NU and 1DL-NU we optimize (\ref{eq:MMSE}) and (\ref{Eq:Distortion_measure}) over $\Delta$ and $\Delta_v$ as well as $a$. As it can be seen in Fig. \ref{Fig_figure1} and \ref{Fig_figure2}, non-uniform quantization improves the performance of both the ICO and 1DL schemes. 1DL-NU outperforms all the other schemes proposed in Section \ref{Sec:achievable transmission schemes} in both the low and high SNR regimes. We expect that SLB is loose in general (especially in the high interference regime), and identifying a tighter lower bound will be instrumental in characterizing the performance limits in this problem. We see that, as expected, NOE outperforms all other encoding schemes, but this is at the expense of a much longer computation time. We also observe that the proposed low-complexity parameterized encoding schemes perform close to NOE, particularly in the high SNR regime.

We also observe that, 1DL outperforms ICO, even though the performances of the two schemes have relatively similar behaviour. Also, in the low interference regime ICA outperforms both ICO and ICO-NU. In Fig. \ref{Fig_figure3}, the size of optimal $\Delta$ versus different channel power constraints for ICO is shown for different values of $\Delta_v$. It can be seen from the figure that by increasing either $P$ or $\Delta_v$, size of $\Delta$ grows non-linearly. This can be easily verified from (\ref{Eq:1}), (\ref{Eq:14}) and (\ref{Eq:15}). Note that (\ref{Eq:15}) tends to $\sigma_s^2$ as $\Delta$ increases. On the other hand, in (\ref{Eq:14}) it can be shown that as $\Delta$ increases, $\Delta_v$ increases too (for a fixed value of $\sigma_T^2$). For high values of $\Delta$ and $\Delta_v$ (\ref{Eq:1}) simplifies to $\Delta=\Delta_v\sqrt{P-\sigma_s^2}$. This linear relation is also observed in the figure. In Fig. \ref{Fig_figure4} the size of the quantization interval $\Delta$ for both ICO and 1DL is plotted against the power constraint. As it is seen in this figure, for all power constraint values, 1DL uses a larger $\Delta$ than ICO for quantization, which explains the improved performance of 1DL compared to ICO (larger $\Delta$ means that the source is mapped into a larger interval, and hence, can be reconstructed with a smaller average distortion). A similar observation applies also to the ICO-NU and 1DL-NU schemes, and the latter outperforms the former.
%----------------------------------------------
\begin{figure}
\begin{centering}
\includegraphics[scale=0.5]{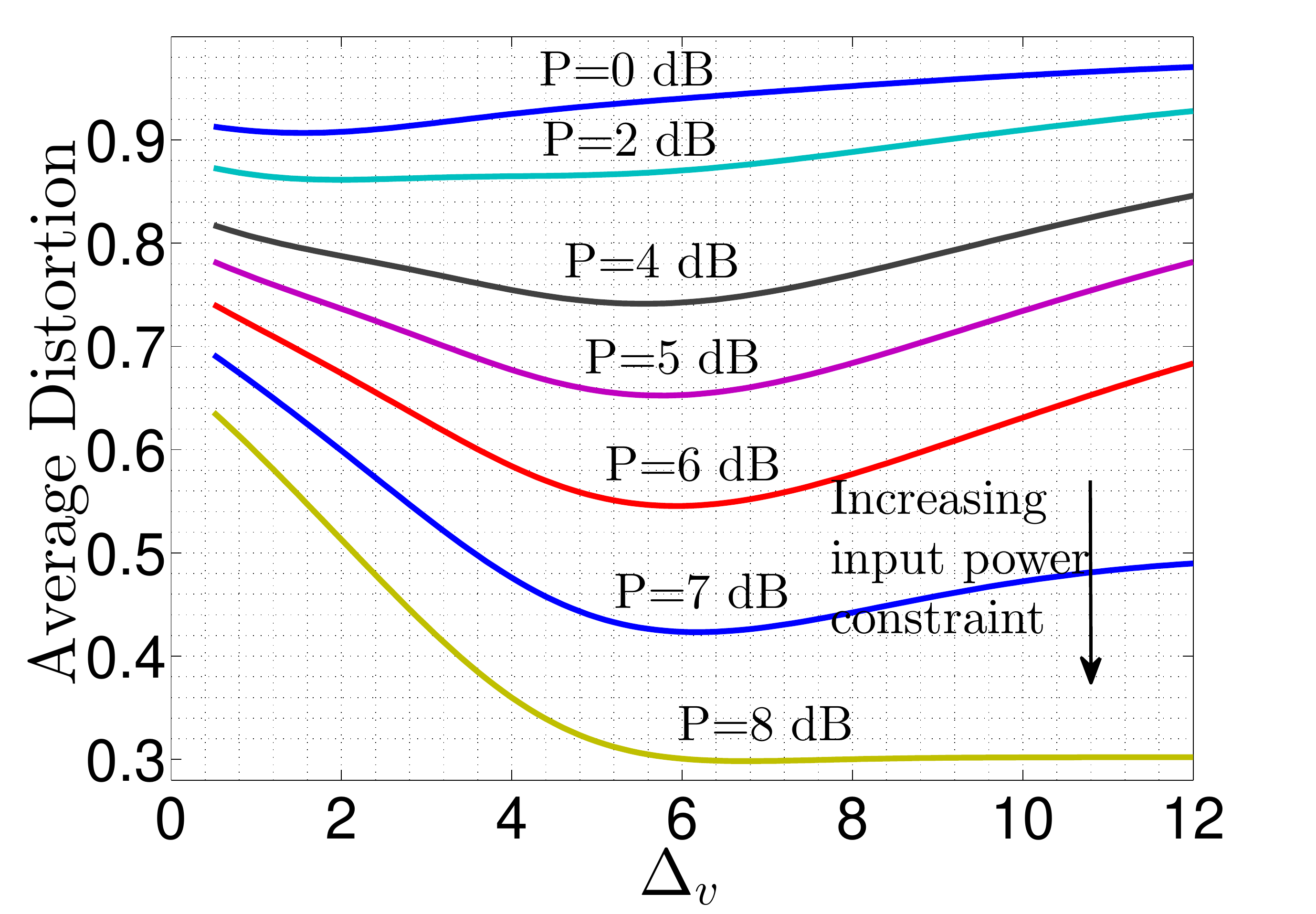}
\par\end{centering}
\caption{\label{Fig_figure5} Average MSE distortion vs. $\Delta_v$ for the ICO for $\sigma_s^2=25$, $\sigma_n^2=1$ .}
\vspace{0mm}
\end{figure}
%----------------------------------------------

%----------------------------------------------
\begin{figure}
\begin{centering}
\includegraphics[scale=0.3]{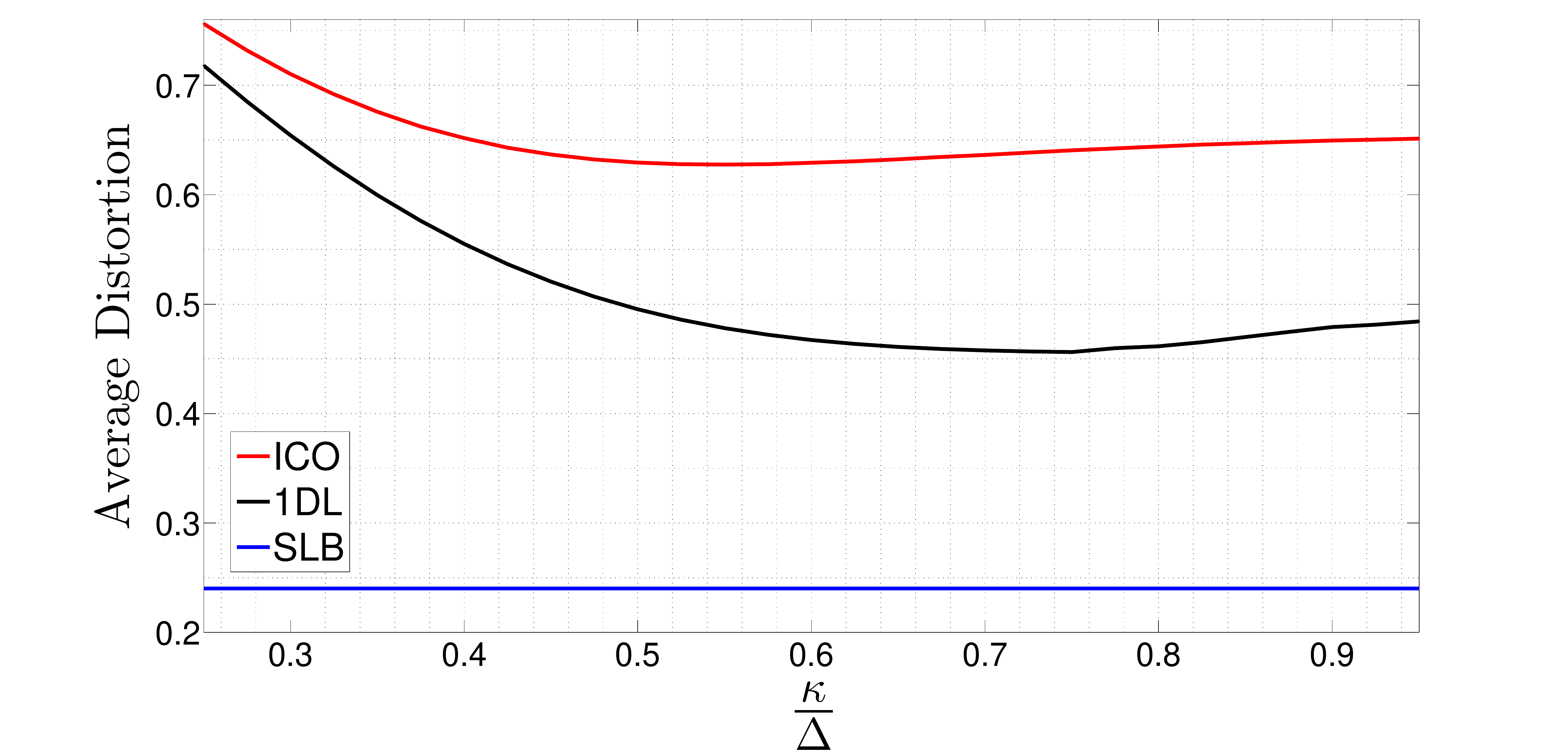}
\par\end{centering}
\caption{\label{Fig_figure6} Average MSE distortion vs. $\frac{\kappa}{\Delta}$ for ICO and 1DL for $P=5$ (dB), $\sigma_s^2=4$, $\sigma_n^2=1$ .}
\vspace{0mm}
\end{figure}
%----------------------------------------------
In Fig. \ref{Fig_figure5} the average distortion versus $\Delta_v$ is plotted for ICO, and for different input power constraints. We observe from the figure that the average  distortion is a convex function of $\Delta_v$. For high power constraints, distortion is almost constant beyond a certain value for $\Delta_v$ (the bottom curve in Fig. \ref{Fig_figure5}). As it can be seen, the higher the input power constraint (since $\sigma_n^2=1$, increasing $P$ is equivalent to increasing SNR) the higher the optimal value for $\Delta_v$, which achieves the minimum average distortion.

In Fig. \ref{Fig_figure6} the average distortion with respect to normalized ($\frac{\kappa}{\Delta}=\frac{\Delta-d}{\Delta}$) is plotted for both ICO and 1DL. It is observed that the average  distortion has a minima with regard to the noise gap, d. It is seen from the figure that there is space to improve the achievable average distortion by optimizing over $d$. Since optimizing the achievable distortion over $d,~\Delta_v,~\Delta$ is demanding, we have obtained the noise gap effect d on the final distortion only for $P=5~\text{dB}$ in Fig \ref{Fig_figure6} (for the remainder of the simulations we have assumed $d=0$).

\section{Conclusions}\label{Sec:conclusion}

In this paper we have studied the problem of zero-delay transmission of a Gaussian source over an AWGN channel in the presence of known interference at the transmitter. Due to the zero-delay constraint and the memoryless nature of the source samples and the interference signals over time, causal and non-causal availability of the interference information are equivalent in this setting. We have proposed one linear and five non-linear zero-delay JSCC schemes. The linear scheme is based on interference cancellation, whereas the non-linear schemes shape the interference and convert it into structured interference, and use companding for the transmission of the source samples.

We have shown that the proposed non-linear coding schemes can achieve zero-distortion in the limit of zero noise, whereas this is not possible through the linear ICA scheme when the interference is strong. We have also introduced the novel idea of non-uniform interference quantization for this problem, and have shown that the corresponding 1DL-NU scheme achieves the best performance among the proposed parametric transmission techniques.

We have also studied the necessary condition for optimality, and proposed a numerically optimized encoder (NOE) obtained using this optimality condition. While NOE outperforms other proposed encoders, it has a significantly higher computational complexity compared to the parameterized schemes. Based on the numerical results it is shown that 1DL-NU performs closer (among the proposed parameterized schemes) to NOE. We have also observed that the structure of the encoder mapping of the proposed parameterized transmission schemes, resemble that of the encoder mapping obtained numerically in the NOE scheme. Based on our numerical performance results and the latter observation, we argue that the proposed low-complexity parameterized transmission schemes can be instrumental in practical systems to achieve reasonably good performance with limited computational resources.

\bibliographystyle{ieeetran}
\bibliography{ref}

\end{document}